\documentclass[journal]{IEEEtran}

\ifCLASSINFOpdf
\else
\fi
\usepackage{amsmath}
\usepackage{algorithmic}
\usepackage[T1]{fontenc}
\usepackage{array}
\usepackage{fixltx2e}
\usepackage{url}
\usepackage{cite}
\usepackage{graphicx}
\usepackage{float}
\usepackage{verbatim}
\usepackage[boxruled,linesnumbered]{algorithm2e}
\usepackage{breqn}

\begin{document}

\title{Extraction of Unaliased High-Frequency Micro-Doppler Signature using FMCW Radar}

\author{Soorya~Peter, and V.V~Reddy,~\IEEEmembership{Member,~IEEE.}

\thanks{Soorya and Vinod are with IIITB, Bengaluru-560078, India. e-mail: (soorya.peter@iiitb.ac.in, vinod.reddy@iiitb.ac.in).}
\thanks{Manuscript received March 01, 2022.}}

\markboth{Journal of \LaTeX\ Class Files,~Vol.~14, No.~8, August~2015}%
{Shell \MakeLowercase{\textit{et al.}}: Bare Demo of IEEEtran.cls for IEEE Journals}

\maketitle

\begin{abstract}
Micro-Doppler signature is a potent feature that has been used for target identification and micro-motion parameter estimation. The extraction of high frequency micro-Doppler signature from frequency modulated continuous wave (FMCW) radar along with the target range and velocity is the problem considered in this article. The severe aliasing of the high micro-Doppler frequency spread is circumvented by the fast time processing in the proposed method. The use of range-Doppler (RD) filtering and empirical mode decomposition (EMD) enables effective out-of-band and in-band noise suppression. Simulation studies and experimental results present the effectiveness of the proposed approach.\\

\end{abstract}


\section{Introduction}

\IEEEPARstart{T}{he} unique micro-motions generated within each target enables target identification and characterization using radar. These micro-motions manifest themselves in the target response as modulated Doppler signature \cite{2003Chen,2011Chen,2006Chen}. This modulation is known to depend on both target characteristics as well as the characteristics of the radar being employed. Since the micro-motions vary with time due to target dynamics, micro-Doppler signatures are often studied with joint time-frequency transforms \cite{2000Chen}, \cite{1998Chen}, exposition of which enables target identification, classification and parameter estimation \cite{2010Liu}, \cite{2014Chen}. In particular, Short-time Fourier transform or Gabor transform is commonly used \cite{2017 Rahman}. Due to the trade-off that exists between time and frequency resolution in such techniques, wavelet decomposition and joint time-frequency distributions have been employed in the literature~\cite{2017 Rahman, 2003Chen}. 

Classification of targets based on micro-Doppler signature has benefited several applications \cite{2018Baptist,2014Villeval,2016Ren,2009Kim,2014Rachel}. Traditionally, micro-Doppler signature of airborne targets has been studied extensively for aircraft or helicopter identification \cite{2006Lin}, \cite{2014Molchanov}. Helicopter parameters such as the length of blade, number of blades, and angular blade rotation rate are estimated from the signature to precisely identify the helicopter model~\cite{2008Cilliers, 2019Singh, 2014Rachel}. More recently, micro-Doppler signature is studied in different contexts including the identification of people based on their unique signature generated by their gait characteristics \cite{2018Baptist} and discrimination between pedestrians, vehicles, and animals in automotive context\cite{2014Villeval}, to mention a few. Micro-Doppler signature has also been employed in applications such as vital signs monitoring and human activity detection \cite{2016Ren},\cite{2009Kim}. 

With substantial increase in the use of UAV for variety of applications, intrusion of UAV into \textit{no flight zone} is a cause of concern. The small radar cross section (RCS), high speed and easy maneuver of UAV add to its stealth capabilities. Detection and characterization of drones is therefore extremely important and challenging. Micro-Doppler signature generated by UAV has been studied for identifying and characterizing them in \cite{2017Jian, 2017 Duncan}

The signatures of drones and birds received by radar are studied to have similar characteristics such as the RCS and velocity profiles in \cite{2018 Rahman, 2014Harmanny, 2021Tahmoush}. As a consequence, UAV detection systems exhibit increased false alarm rate. In order to improve the detection performance, micro-Doppler signatures generated by these targets are studied \cite{2016 Ritchie} and exploited for classification \cite{2019 Bjorklund},\cite{2019 Aldowesh}. Micro-Doppler signatures are shown to enable identification of the model of the UAV in~\cite{2017 Fuhrmann},\cite{2019 Bjorklund}. Convolutional Neural network (CNN) has been employed in \cite{2017 Kim} for UAV classification. The micro-Doppler signature and its frequency domain representation are merged to form a Doppler image that is provided to the CNN for classification. 

Micro-Doppler signature is often characterized by time-varying frequency spread that is modulated over the target Doppler frequency. In applications such as human activity detection, vital signs monitoring and classification of automotive targets, the frequency spread due to micro-Doppler signature is lower than the entire Doppler range. This enables undistorted micro-Doppler signal analysis along the slow-time of pulsed-wave (PW) radar \cite{2016 Ritchie}. Continuous-wave (CW) radar, on the other hand, is capable of handling significantly large micro-Doppler frequency spread, such as in the ca the use of CW radar in such applications. 

The use of frequency modulation is widely known to enable CW radar to estimate target range. However, the capability of the frequency modulated CW (FMCW) radar to process micro-Doppler signature is similar to that of PW radar since Doppler processing is performed along the slow-time. For targets with large micro-Doppler frequency spread, such as the UAVs, slow-time signal analysis is adversely affected by severe aliasing. In \cite{2021Peter}\cite{2021Vinod}, an approach was presented to extract micro-Doppler signature by stitching the down-converted receive signal from multiple chirps as a continuous signal that preserves high-frequency micro-Doppler signature. With the sampling frequency along the fast-time higher than the micro-Doppler frequency spread, alias-free analysis can be performed.


In this work, we extend the work presented in \cite{2021Peter,2021Vinod} by first studying the FMCW radar signal model with the high frequency micro-Doppler signature embedded within. We then analyze the extracted micro-Doppler signature by the slow-time Doppler processing and the technique presented in \cite{2021Peter} to identify some limitations. The extent of frequency spread due to micro-Doppler signature induced by the blade rotation is studied to gain insights into the Doppler and Range spread observed in the Range-Doppler map. Using this understanding, we then propose a technique to extract the high frequency micro-Doppler signature. The proposed technique also addresses the limitations of the technique presented in \cite{2021Peter} by employing empirical mode decomposition (EMD) before the time-frequency analysis of specific modes that preserve the micro-Doppler signature.


The outline of this article is as follows: In Section II we present the signal model for FMCW radar in the presence of a target with micro-motions. We then review micro-Doppler extraction process along slow-time followed by a review of the technique in~\cite{2021Peter}\cite{2021Vinod} for high frequency micro-Doppler extraction. The limitations of these techniques that are identified are addressed in Section III where we present an enhanced EMD-based technique to extract high frequency micro-Doppler signature effectively. Simulation studies are presented in Section IV to evaluate the effectiveness of the proposed technique in adverse signal-to-noise ratio conditions. In Section V, we present the micro-Doppler signature extracted by the proposed technique for a hovering UAV. 

\section{Problem definition and Review of micro-Doppler signature extraction }

\subsection{FMCW radar signal model in the presence of micro-motions}
Consider FMCW radar operating from a start frequency of $f_0$ and chirprate of $\mu$ over a chirp duration of $T_{\mathrm{c}}$. In the presence of $K$ targets within the field-of-view of the radar, the deramped signal is given by
\begin{align}
\label{IFsignal}
   y^{(l)}(t) &= \sum_{k=1}^K \alpha_k e^{-\jmath\big\{2\pi f_0 \tau_k + 2\pi \mu \tau_k t - \pi \mu \tau_k^2\big\}} + n^{(l)}(t) & &\\
   & \approx  \sum_{k=1}^K \alpha_k e^{-\jmath\big\{2\pi f_0 \tau_k + 2\pi \mu \tau_k t\big\}} + n^{(l)}(t)  & &
   \end{align}
where $\alpha_k$ is the received signal strength from the $k$th target, and $n^{(l)}(t)$ is the additive noise signal for the $l$th chirp response. The delay for $k$th target, parameterized by range $R_k$ and relative radial velocity $v_k$, is given by
\begin{equation}
\tau_k = \frac{2}{c}\big[R_k + v_k\tilde{t}\big],
\end{equation}
where $\tilde{t}=t+(l-1)T_{\mathrm{CRI}}$, $t\in (0,T_{\mathrm{c}})$, $T_{\mathrm{CRI}}\geq T_{\mathrm{c}}$ is the chirp repetition interval, and $c$ is the propagation speed.

Since our focus is on the study of target micro-Doppler signature, we consider only one target having a rotor with $K_B$ blades, each of length $L_B$. The target is at a distance of $R_0$ from the radar and has a relative radial velocity of $v_0$. The received signal after down conversion that constitutes the response from the target body and the blades is given by
\begin{equation}
\label{SM_md}
    y^{(l)}(t) = {\alpha^{body}}e^{\jmath\Big\{\Big(\frac{4\pi}{\lambda}+\frac{4\pi\mu{t}}{c}\Big)\Big(R_0+v_0\tilde{t}\Big)\Big\}}+\sum_{b=1}^{K_B}s_{b}^{(l)}(t)+n^{(l)}(t).
\end{equation}
where $s_{b}^{(l)}(t)$ is the response due to $b$th blade. The signal scattered by a scatterer  at a distance of $l_b$ on the blade will introduce a delay of 
\begin{equation}
\tau_k \approx \frac{2}{c}\Big[R_k + v_k\tilde{t}+l_p \cos\beta \cos{(\omega_b(\tilde{t}))}\Big].
\end{equation}
Here, $\beta$ is the elevation angle from radar to the target and $\omega _b(\tilde{t}) = \Omega \tilde{t}+\psi_b$ is a function of the blade rotation rate $\Omega$ and initial blade offset angle $\psi_b$. Integrating the response due to the entire blade length $L_B$, the blade response can be evaluated as
\begin{equation}
\begin{split}
\label{IF_sig_one_blade}
 s_{b}^{(l)}(t) = L_B\alpha_b e^{\jmath\Big\{\Big(\frac{4\pi}{\lambda}+\frac{4\pi\mu{t}}{c}\Big)\Big(R_0+v_0\tilde{t}+{\frac{L_B}{2}}\cos{\beta}\cos{\omega_b(\tilde{t})}\Big)\Big\}} \\ \mathrm{sinc}\Big(\Big(\frac{4\pi}{\lambda}+\frac{4\pi\mu t}{c}\Big)\Big(\frac{L_B}{2}\cos{\beta}\cos{\omega_b(\tilde{t})}\Big)\Big),
\end{split}
\end{equation}
where $\alpha_b$ is the received signal strength from $b$th blade. We note that the blade response is a frequency modulated sinc function whose instantaneous frequency can be obtained as 
\begin{equation}
\begin{split}
\label{mdfreq}
       f_{mD}(t) = f_{R_0} + f_{D_0} +\frac{2L_B\mu{\cos{\beta}}}{c}\Big(\cos{\omega _b(\tilde{t})}-{t}\Omega\sin{\omega _b(\tilde{t})}\Big)\\+\frac{2\mu}{c}v\tilde{t}-\frac{2L_B\Omega{\cos{\beta}}}{\lambda}\Big(\sin{\omega _b(\tilde{t})}\Big),
\end{split}
\end{equation}
where $f_{R_0}=\frac{2\mu}{c} R_0 $  and $f_{D_0}=\frac{2v_0}{c} f_0$. We observe that the instantaneous frequency is a function of the chirp-rate, blade length, blade rotation rate, wavelength, and the target elevation angle. Due to the blade rotation angle, the instantaneous frequency spreads on either sides of the target Doppler frequency. It is evident that the frequency spread increases with larger chirp-rate and radar operating frequency.

In order to study the variations in frequency with time, joint time-frequency analysis is performed. However, the study of micro-Doppler signature is affected by undesired response from other targets in the scene. In what follows, we review the conventional technique that is widely used for micro-Doppler signature extraction in FMCW radar.

\subsection{Review of Slow-Time Micro-Doppler Signature Extraction (STMDSE)}
Unlike the monotonic CW radar, FMCW radar transmits multiple chirps to receive response from the targets. The target Doppler frequency, and therefore the relative radial velocity, is estimated across multiple chirp responses within the Coherent Processing Interval (CPI), similar to PW radar.

\begin{figure}[ht]
    \centering
    \hspace*{-0.5in}
    \includegraphics[width=\linewidth]{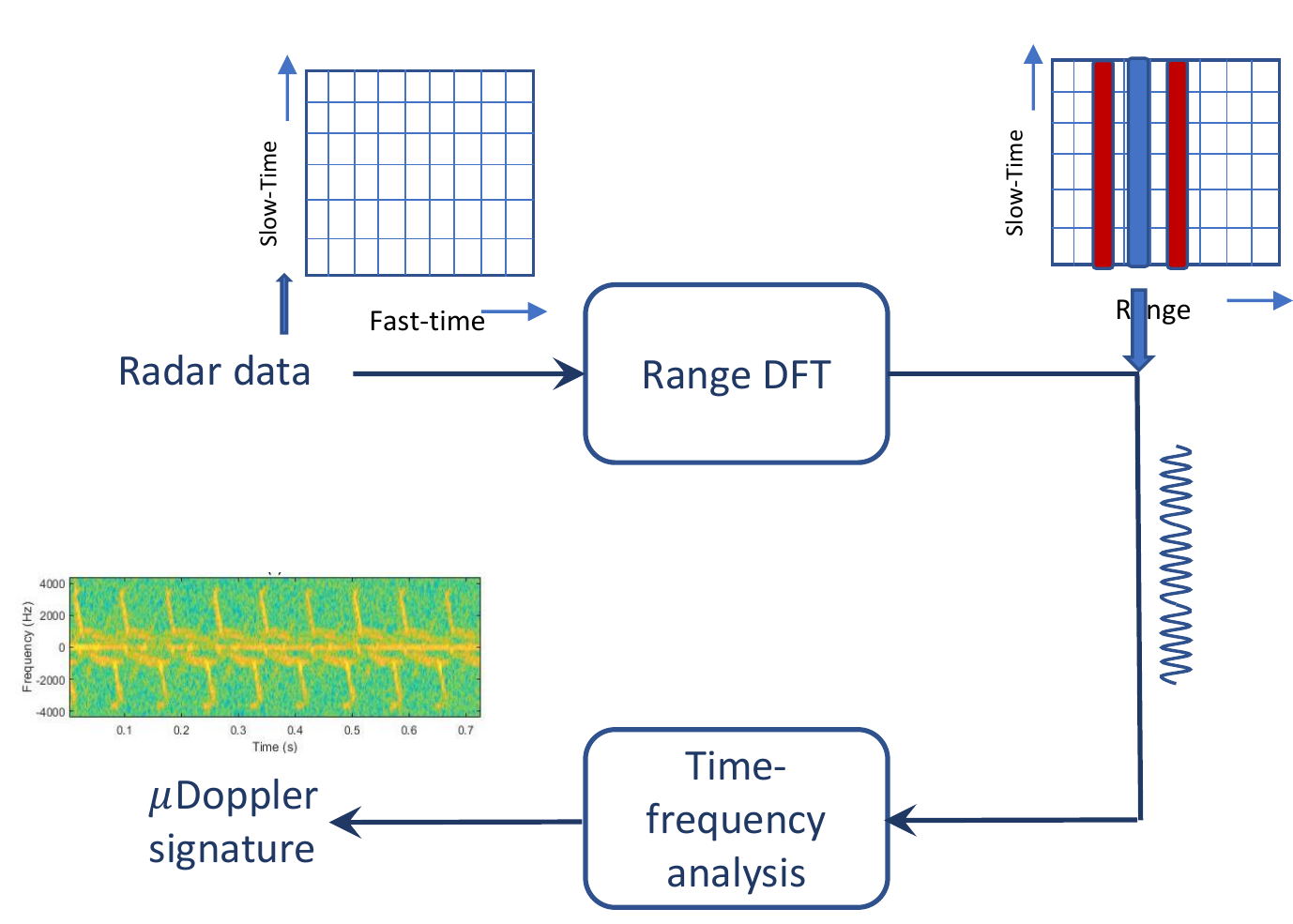}
    \caption{Extraction of micro-Doppler signature using conventional approach.}
    \label{fig:slow-time_BD}
\end{figure}
As observed from (\ref{mdfreq}), the instantaneous frequency is centered around $f_R + f_D$. For $f_R >> f_D$, the Fourier transform along the fast time will be centered at the range bin corresponding to $f_R$. Since the micro-Doppler signature rides on the target Doppler signature, most often, micro-Doppler signature is studied using time-frequency analysis at this range bin across multiple chirp responses (slow time) as shown in Fig.~\ref{fig:slow-time_BD}. This signal along the slow time is inherently sampled at the chirp repetition frequency, $f_{\mathrm{s-ST}}=f_{\mathrm{crf}}$, and therefore introduces an upper limit on the alias-free signal frequency of 
 \begin{equation}
 \label{sampling}
     f_{D_{\mathrm{max}}}+\Delta f_{\mathrm{mD_{max}}} \leq \frac{f_{\mathrm{s-ST}}}{2} ,
 \end{equation}
where $f_{D_{\mathrm{max}}}$ is the maximum Doppler frequency and $\Delta f_{\mathrm{mD_{max}}}$ is the maximum expected frequency deviation due to micro-Doppler. 

\begin{figure}[ht]
    \centering
    \hspace*{0 in}
    \includegraphics[width=\linewidth]{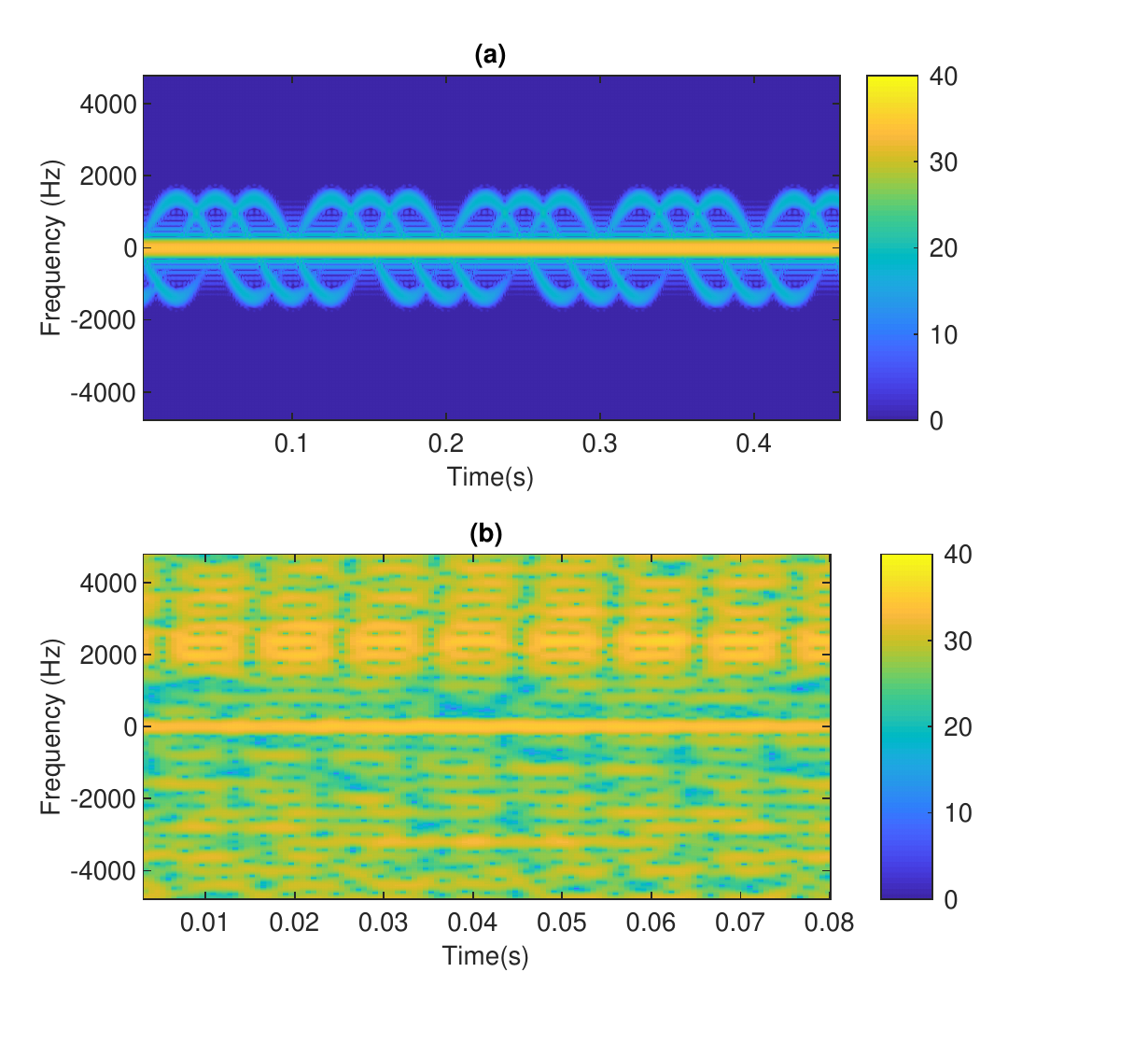}
    \caption{Spectrogram obtained using conventional approach for a target with blades rotating at (a) 600rpm (b) 6000rpm.}
    \label{fig:slow_time_mD}
\end{figure}

This approach is suitable for applications where the micro-Doppler frequency spread is small. However, in applications such as UAV detection where the rotor blades can rotate from 6000-40000rpm, micro-Doppler frequency spread will be significantly large to introduce severe aliasing. We illustrate this aliasing phenomenon for a UAV with rotor blades rotating at 600rpm and 6000rpm in the FOV of FMCW radar operating at 77GHz with a chirp-rate of $10Mhz/\mu{s}$ and $f_{\mathrm{crf}} = 9.57kHz$. The spectrogram of the slow-time signal for 600rpm and 6000rpm cases are plotted in Fig.~\ref{fig:slow_time_mD}.
It is evident from the two subplots that a large rotation rate results in severe aliasing. Extraction of micro-Doppler signature and target characterization is therefore not possible with this approach. 

In order to overcome this limitation, we proposed a novel approach to extract high frequency micro-Doppler signature in \cite{2021Peter,2021Vinod} that is not affected by aliasing. This method will be discussed next.



\subsection{Review of Fast-Time Micro-Doppler Signature Extraction (FTMDSE)}
As opposed to the short-lived pulse in the PW radar, the duty cycle of the transmit chirp in FMCW radar is close to $100\%$. In FTMDSE, the responses for multiple transmitted chirps are appended to form a continuous down-converted signal similar to a CW radar except for a small idle time, $T_{\mathrm{idle}}$ between consecutive chirps. It is important to emphasize that the appended signal is sampled with the fast-time sampling frequency $f_{s-\mathrm{FT}}$ which is usually designed based on the maximum target range, $f_{s-\mathrm{FT}}\geq f_{R_{\mathrm{max}}}$. The target response is embedded in this signal as a monotonic signal with frequency equal to $f_R+f_D$. The micro-Doppler signature will modulate the frequency of this signal as described in (\ref{mdfreq}). 

By choosing the fast-time sampling frequency such that
\begin{equation}
f_{s-\mathrm{FT}} > f_{R_{\mathrm{max}}} + f_{\mathrm{mD-max}},
\end{equation}
it is assured that the signal will embed the micro-Doppler signature without any aliasing. In contrast to STMDSE, FTMDSE method extracts the high-frequency micro-Doppler signature from the fast-time signal.

\begin{figure}[H]
    \centering
    \includegraphics[width=\linewidth]{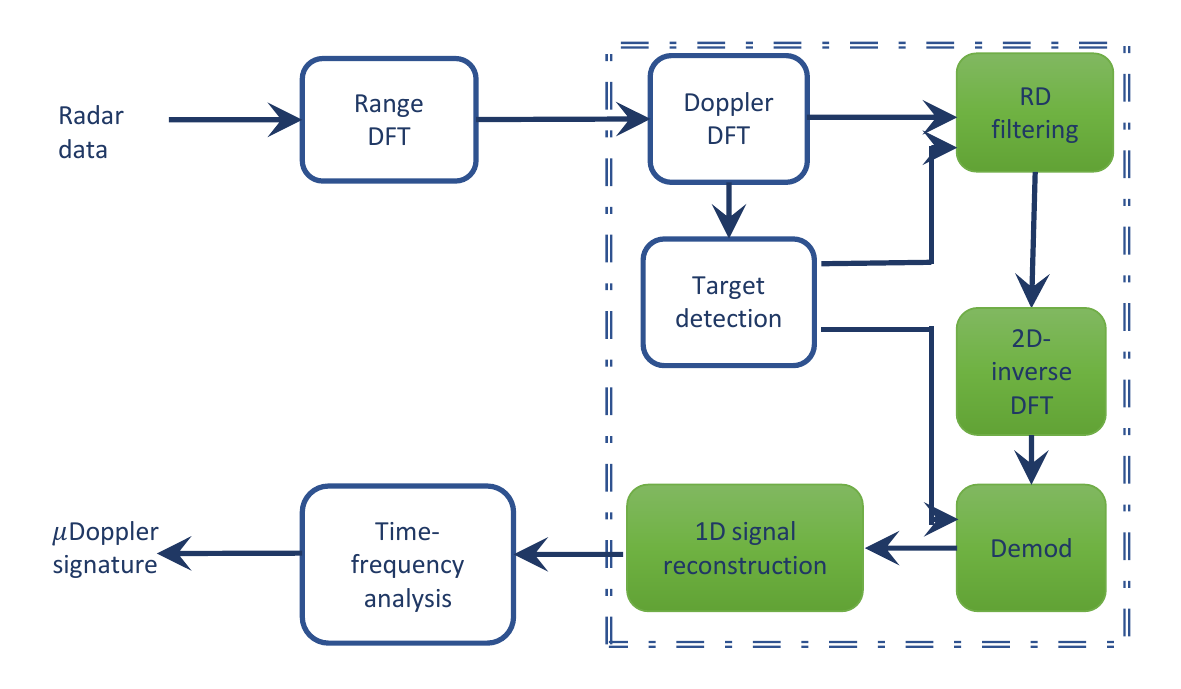}
    \caption{Block diagram of FTMDSE method.}
    \label{fig:BD_FTMDSE}
\end{figure}
\begin{figure}[H]
   \centering
   \hspace*{0in}
    \includegraphics[width=9cm,height=6cm]{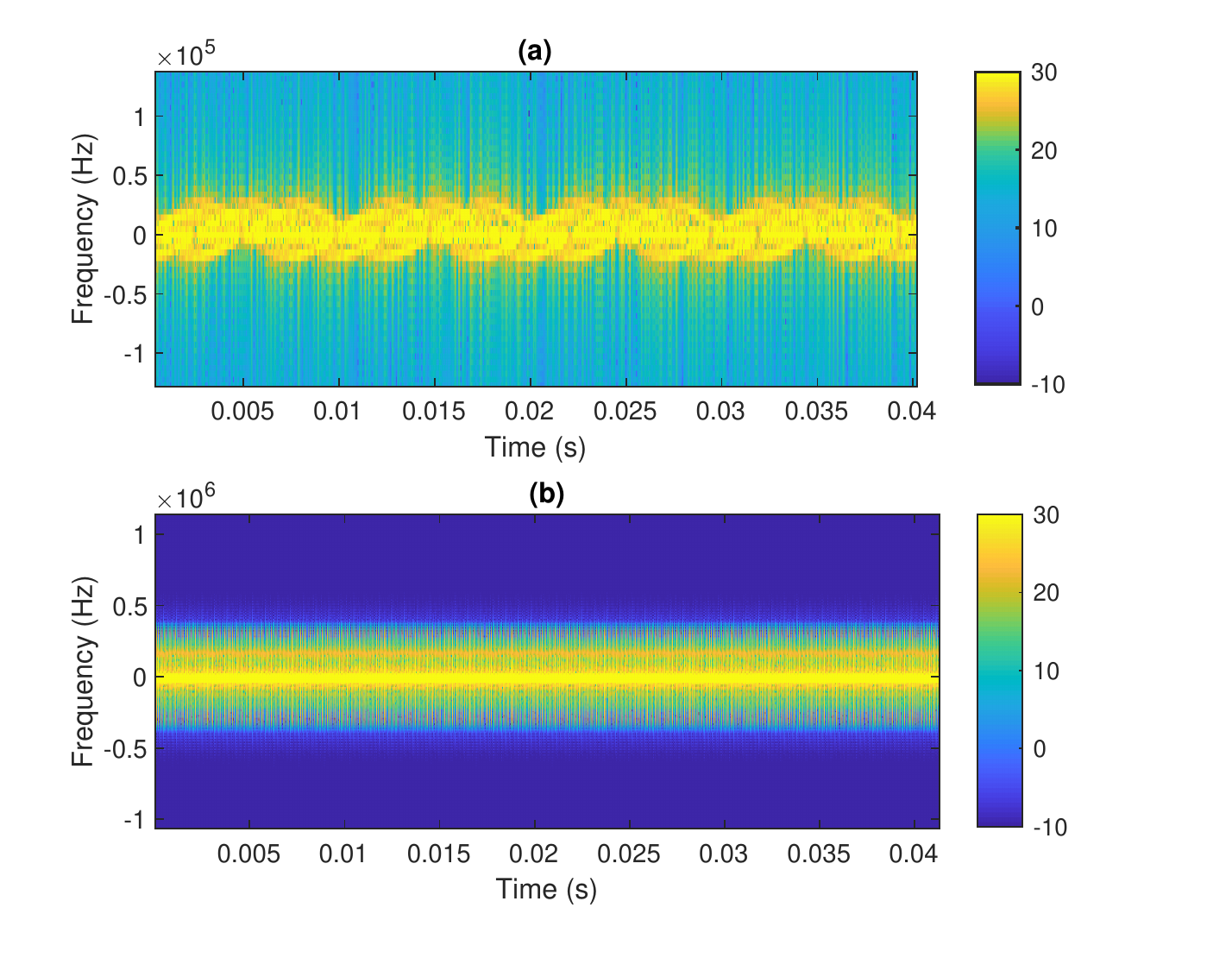}
    \vspace*{-8mm}
   \caption{Spectrogram obtained using FTMDSE method for a (a) target with blades rotating at 6000rpm and (b) stationary target.}
    \label{fig:stationary_target}
\end{figure}

Fig.~\ref{fig:BD_FTMDSE} shows all steps involved in FTMDSE method. The $L$ down-converted chirp responses within the CPI are first subjected to 2D Fourier transform to obtain  
\begin{equation}
    Y(f_R,f_D) = \sum_{n=0}^{N-1}\sum_{l=0}^{L-1}y^{(l)}(nT_s)e^{\jmath\{2\pi(nT_sf_R+f_DlT_{cri})\}}
\end{equation}
where $N$ is the number of fast-time samples available per chirp. Among the targets detected in the Range-Doppler (RD) map, the target to be verified for its micro-Doppler signature is identified by its estimated range and velocity $(\hat{R}_0, \hat{v}_0)$. A range-Doppler filter, $G(f_{R},f_D)$, that is centered around $(\hat{R}_0, \hat{v}_0)$ is employed to retain only the target response

\begin{equation}
     Y_{\mathrm{filt}}(f_R,f_D) = Y(f_R,f_D) G(f_R,f_D).
\end{equation}
The 2D inverse Fourier transform of $Y_{\mathrm{filt}}(f_R,f_D)$ provides the filtered time-domain signal $y_{\mathrm{filt}}^{(l)}(nT_s)$. With the estimated range and velocity, this signal is then demodulated to obtain
\begin{equation}
\begin{split}
    \widetilde{y}_{\mathrm{filt}}^{(l)}(nT_s) = y_{\mathrm{filt}}^{(l)}(nT_s)e^{\jmath\Big\{\Big(\frac{4\pi}{\lambda}+\frac{4\pi\mu{t}}{c}\Big)\Big(R_0+v_0\tilde{t}\Big)\Big\}} .
\end{split}
\end{equation}
The resultant signal preserves only the target micro-Doppler signature. Defining the row vector $\mathbf{\tilde{y}}_{\mathrm{filt}}^{(l)}=[\widetilde{y}_{\mathrm{filt}}^{(l)}(0)\;\ldots\;\widetilde{y}_{\mathrm{filt}}^{(l)}((N-1)T_s)]$, \cite{2021Peter} constructs the signal  
\begin{equation}
\label{appended_signal}
    \widetilde{y}_{\mathrm{r}}(nT_s) = [\mathbf{\tilde{y}}_{\mathrm{filt}}^{(0)},\mathbf{\tilde{y}}_{\mathrm{filt}}^{(1)},\;\ldots\;,\mathbf{\tilde{y}}_{\mathrm{filt}}^{(L-1)}].
\end{equation}
The discontinuities that arise between two signals are filtered using a low pass filter before time frequency analysis of $\widetilde{y}_{\mathrm{r}}(nT_s)$. The micro-Doppler signature obtained by this approach for a target with a blade rotation rate of $6000~$rpm is shown in Fig.~\ref{fig:stationary_target}(a). While we observe that the technique extracts high frequency micro-Doppler signature without aliasing, there are some limitations.

Practical FMCW radars are known to have an idle time during which the synthesizer resets the transmit frequency back to the start frequency. Ignoring this idle time will introduce error in micro-motion parameters that are estimated from the signature. Furthermore, the discontinuity between the chirp responses introduces periodic frequency spread in the spectrogram with a periodicity of $T_{\mathrm{CRI}}$. Alternatively, interpolation using  Autoregressive models between every consecutive fast-time signals will be inaccurate and computationally expensive. 

The use of low pass filter in \cite{2021Peter} with predetermined cutoff frequency results in the suppression of out-of-band noise and emphasizing the in-band noise even in the absence of any micro-Doppler signature as shown in Fig.~\ref{fig:stationary_target}(b).
This can lead to misleading interpretation of the signal and prompts alternate mechanism to suppress discontinuities between chirp responses. In the next section, we propose an EMD-based technique to address these two set backs of FTMDSE method.


\section{Proposed EMD-based FTMDSE method}
\begin{figure}[H]
    \centering
    \includegraphics[width=\linewidth]{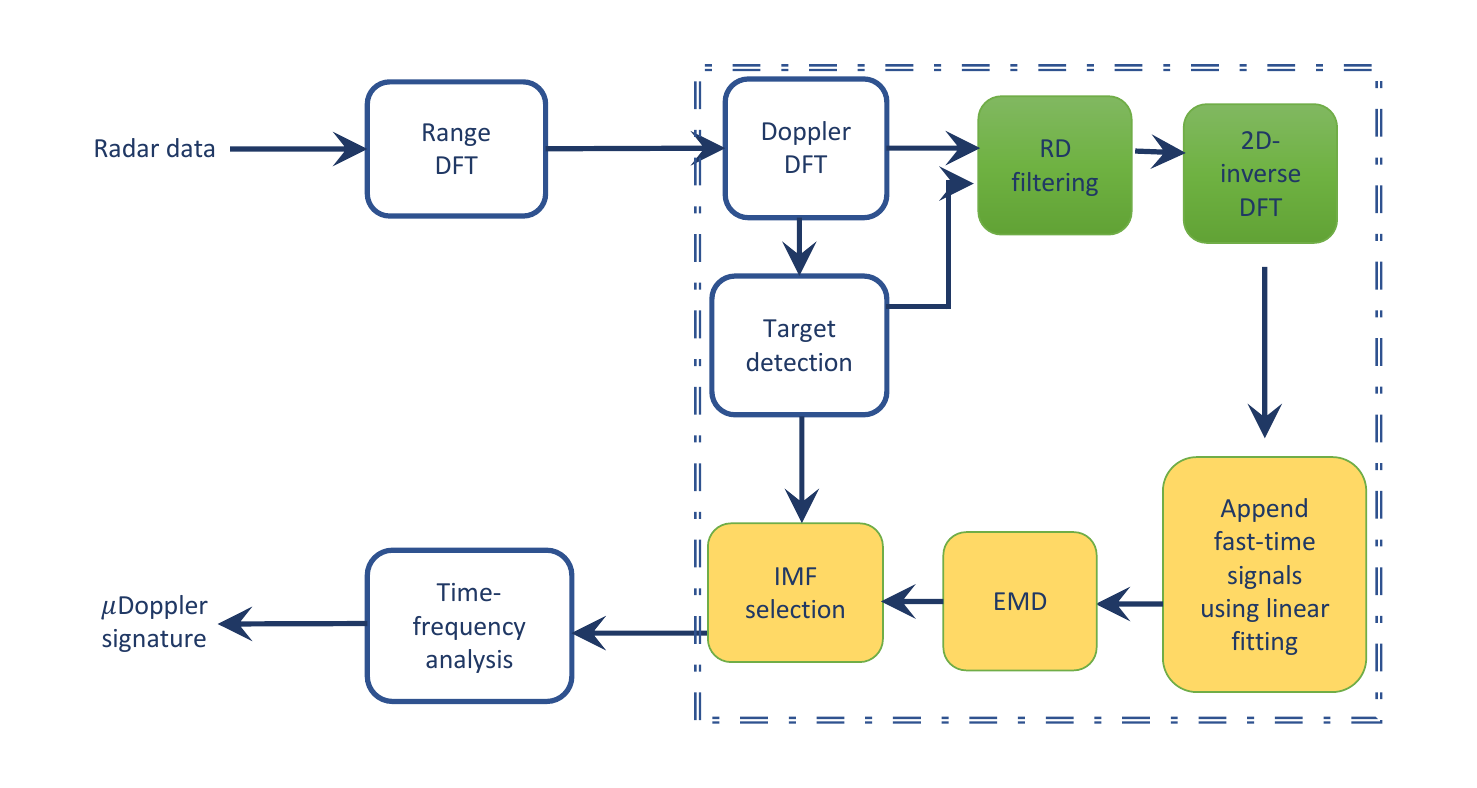}
    \caption{Block diagram of the proposed EMD-based FTMDSE method.}
    \label{fig:updated_BD}
\end{figure}

We now propose a technique that adopts the FTMDSE approach and addresses the limitations of the earlier presented technique~\cite{2021Peter}. The block diagram for the proposed technique is shown in Fig.~\ref{fig:updated_BD} to have some steps to substitute the low pass filter employed in FTMDSE method. Details of the variations will also be discussed.

Consider the signal received by the radar due to a single target with micro-motions to abide by the model defined in (\ref{SM_md}). The target response over the entire $L$ chirps within the CPI is first subjected to the Range DFT. Ignoring the negligible quadratic phase term and the phase term with $(2(\mu/c)t v_0(l-1)T_{\mathrm{CRI}})$, we obtain,
\begin{multline}
\overline{Y}^{(l)}(f_R) = \alpha^{body} \delta(f_R-(f_{R_0}+f_{D_0})) e^{\jmath\Big\{ 2\pi \frac{2v_0 f_0}{c}(l-1)T_{\mathrm{CRI}} +\phi\Big\}} \\  + \sum_{b = 1}^{K_B} \overline{S}_b^{(l)}(f_R) + \overline{N}^{(l)}(f_R), 
\end{multline}
where $\delta(.)$ is the impulse function obtained by the Fourier transform of the complex sinusoid, $\phi=4\pi R_0/\lambda$ is the constant phase term, $\overline{Y}^{(l)}(f_R)$, $\overline{S}_b^{(l)}(f_R)$ and $\overline{N}^{(l)}(f_R)$ are obtained by the 1D Fourier transform of $y^{(l)}(t)$, $s_b^{(l)}(t)$ and $n^{(l)}(t)$, respectively. When the Doppler DFT is applied at each range frequency across all the chirp responses, the RD map is obtained as
\begin{multline}
Y(f_R,f_D) =\alpha^{body} \delta (f_R-(f_{R_0}+f_{D_0})) \delta (f_D-f_{D_0}) e^{\jmath\phi}  \\ + \sum_{b = 1}^{K_B} S_b(f_R,f_D) + N(f_R,f_D), 
\end{multline}
where $f_{D_0}=\frac{2v_0 f_0}{c}$ is the target Doppler frequency. The peak identified in $Y(f_R,f_D)$ provides the target range and Doppler estimates $\hat{f}_{R_0}$ and $\hat{f}_{D_0}$.

\begin{figure}[H]
    \centering
    \includegraphics[width=9cm,height=7cm]{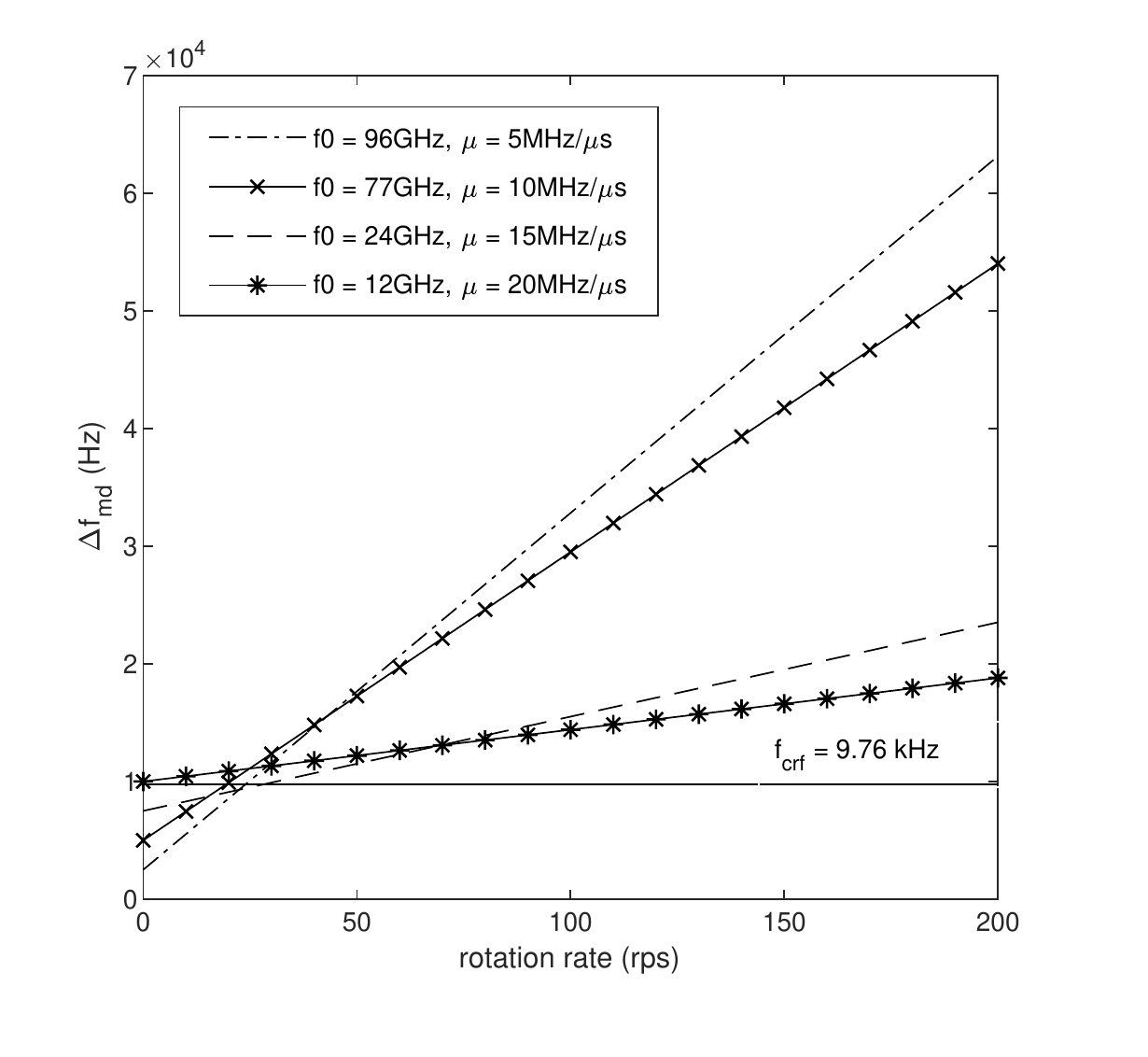}
    \vspace*{-8mm}
    \caption{Variation in $\Delta f_{\mathrm{mD}}$ against blade rotation rate.}
    \label{fig:fmd}
\end{figure}

The response from the blades, $S_b(f_R,f_D)$, will be centered around $\hat{f}_{R_0}+\hat{f}_{D_0}$. However, as observed in (\ref{mdfreq}), the instantaneous frequency of $s_b(t,l)$ varies with time due to the blade rotation, thereby depriving a closed form expression for $S_b(f_R,f_D)$. The maximum frequency spread around $\hat{f}_{R_0}+\hat{f}_{D_0}$ can be obtained as
\begin{equation}
\label{maxMD-Spread}
\Delta f_{\mathrm{mD}} = f_{\mathrm{mD-max}}-f_{\mathrm{mD-min}} = \frac{2L_B\Omega}{c}\{\mu T_c + f_0 \},
\end{equation}
where $f_{\mathrm{mD-max}}$ and $f_{\mathrm{mD-min}}$ are obtained from (\ref{mdfreq}) for $\omega_b(t)= -90$ and $\omega_b(t)= 90^\circ$, respectively. Besides the dependence of $\Delta f_{\mathrm{mD}}$ on the blade rotation rate $\Omega$, it is interesting to note that it is also a function of the blade length, chirprate and operating frequency. In order to understand the sensitivity of $f_{\mathrm{mD}}$ to these parameters, we study the extent of micro-Doppler frequency spread $\Delta f_{\mathrm{mD}}$ against increasing rotation rate for four different radar parameters in Fig.~\ref{fig:fmd}.

We note that $\Delta f_{\mathrm{mD}}$ increases with $\Omega$ for all the radar operating cases. The larger spread in $f_{\mathrm{mD}}$ for higher radar operating parameters, asserts the severity of micro-Doppler signature. For a target with micro-motions such that 
\begin{equation*}
\Delta f_{\mathrm{mD}} > \frac{2\mu}{c}\delta r = \frac{\mu}{B},
\end{equation*}
where $\delta r=c/(2B)$, it is evident that $\overline{S}_b(f_R,l)$ will exhibit a spread in the range frequency. Likewise, when 
\begin{equation*}
f_{\mathrm{crf}}>\Delta f_{\mathrm{mD}} > f_{\mathrm{crf}}/L,
\end{equation*}
$S_b(f_R,f_D)$ exhibits a spread over multiple Doppler bins. However, when 
\begin{equation*}
|f_{D_0}+\Delta f_{\mathrm{mD}}|>f_{\mathrm{crf}},
\end{equation*}
we encounter aliasing in the Doppler dimension. In such a scenario, the study of micro-Doppler using STMDSE method fails due to aliasing. Figure~\ref{fig:fmd} provides us with the insight on the severity of aliasing for $\Delta f_{\mathrm{mD}}$ studied in comparison with $f_{\mathrm{crf}}$. From this study, it is evident that 
\begin{equation}
\Delta f_{\mathrm{mD}}>\max{\Big\{\frac{\mu}{B}, f_{\mathrm{crf}}\Big\}}
\end{equation}
results in a spread along both the range as well as the Doppler domains around the target RD bin.

\subsection{Range-Doppler filtering}
Clutter-suppression filter is commonly used in CW-radar to exclude the response from the background. In FMCW radar, one can suppress undesired targets,  clutter and noise by employing a Range-Doppler filter. The pass band of this filter is centered around the desired target range and Doppler frequencies, thereby suppressing contribution from  different range and Doppler bins. 

The frequency response of a Gaussian filter is given by
\begin{equation}
 \label{filter}
    G(f_R,f_D) = \exp\Bigg(-\Bigg\{\frac{(f_R-\hat{f}_{R_0})^2}{2\sigma_{f_R}^2}+\frac{(f_D-\hat{f}_{D_0})^2}{2\sigma_{f_D}^2}\Bigg\}\Bigg),
\end{equation}
where $\hat{f}_{R_0}$ and $\hat{f}_{D_0}$ are the estimated range and Doppler frequencies of the desired target, $\sigma_{f_R}^2$ and $\sigma_{f_D}^2$ regulate the filter bandwidth in the two dimensions. The 3-dB cut-off frequency of the range filter $f_c$ is related to $\sigma_R$ as $f_c=\sqrt{\log(2)}\sigma_{f_R}$. These parameters are determined usually by the target size and the expected dynamics in the scenario. 

In the presence of micro-Doppler however, the filter bandwidth in the two dimensions have to be chosen based on the frequency spread, $\Delta f_{\mathrm{mD}}$, as discussed above. For $\Delta f_{\mathrm{mD}}>f_{\mathrm{crf}}$, the frequency spread will be observed over the entire Doppler dimension and over neighbouring range bins. It is therefore necessary to have an all pass filter along the Doppler (by setting a large $\sigma_{f_D}$), failing which will result in inaccurate micro-Doppler signature recovery. The bandwidth of the range filter also has to be chosen such that $f_c>\Delta f_{\mathrm{mD}}/2$. We then recover the filtered time-domain signal $y_{\mathrm{filt}}^{(l)}(nT_s)$ by the 2D inverse-DFT of $Y_{\mathrm{filt}}(f_R,f_D)$. 

The failure of STMDSE along the slow-time inspired the use of the signal along the fast time for micro-Doppler signature analysis~\cite{2021Peter}. The response from multiple chirps are appended to obtain a longer duration signal for the analysis. Practical FMCW radar with a chirp duration of $T_c$ however, will have an idle duration $T_{\mathrm{idle}} = T_{\mathrm{CRI}}-T_c$ during which the response is not available. It is important to handle this duration carefully, failing which would result in artifacts that can adulterate the extracted signature. 

\subsection{Linear fitting}
Appending the fast-time signals without accounting for $T_{\mathrm{idle}}$ will have two severe consequences. The discontinuity in time domain between any two fast-time signals will result in an undesired wide frequency spread in the time-frequency analysis. This periodic frequency spread adversely affects the signal interpretation. Directly appending the signature without accounting for $T_{\mathrm{idle}}$ results in inaccurate estimation of target micro-motion parameters such as the periodicity.

We therefore have to interpolate between consecutive signals over a duration of $T_{\mathrm{idle}}$. Since interpolation has to be performed after every chirp response, a computationally efficient method is required that does not affect the time-frequency analysis. We therefore first define the interpolated signal by
\begin{equation}
\widetilde{y}(nT_s) = [\mathbf{y}_{\mathrm{filt}}^{(l)}\; \mathbf{y}_{\mathrm{interp}}^{(l,l+1)}\;\mathbf{y}_{\mathrm{filt}}^{(l+1)}\;\ldots\;\mathbf{y}_{\mathrm{filt}}^{(L-1)}],
\end{equation}
where $\mathbf{y}_{\mathrm{interp}}^{(l,l+1)}=[y_{\mathrm{interp}}^{(l,l+1)}(0),\ldots, y_{\mathrm{interp}}^{(l,l+1)}((Q-1)T_s)]$, with $T_{\mathrm{idle}}=QT_s$, is the interpolated signal obtained as
\begin{multline*}
y_{\mathrm{interp}}^{(l,l+1)}(n)=y_{\mathrm{filt}}^{(l)}((N-1)T_s) + \\\frac{y_{\mathrm{filt}}^{(l+1)}(0)-y_{\mathrm{filt}}^{(l)}((N-1)T_s)}{T_{\mathrm{idle}}}nT_s.
\end{multline*}
Despite linear interpolation being an inaccurate estimate of the signal between the chirp responses, we employ it in order to introduce deterministic low frequency components over this duration in the time-frequency analysis as opposed to a large spread observed due to the discontinuity. The use of EMD subsequently will also have a predictable outcome over the idle time.

\begin{figure}[ht]
    \centering
    \includegraphics[width=9cm,height=7cm]{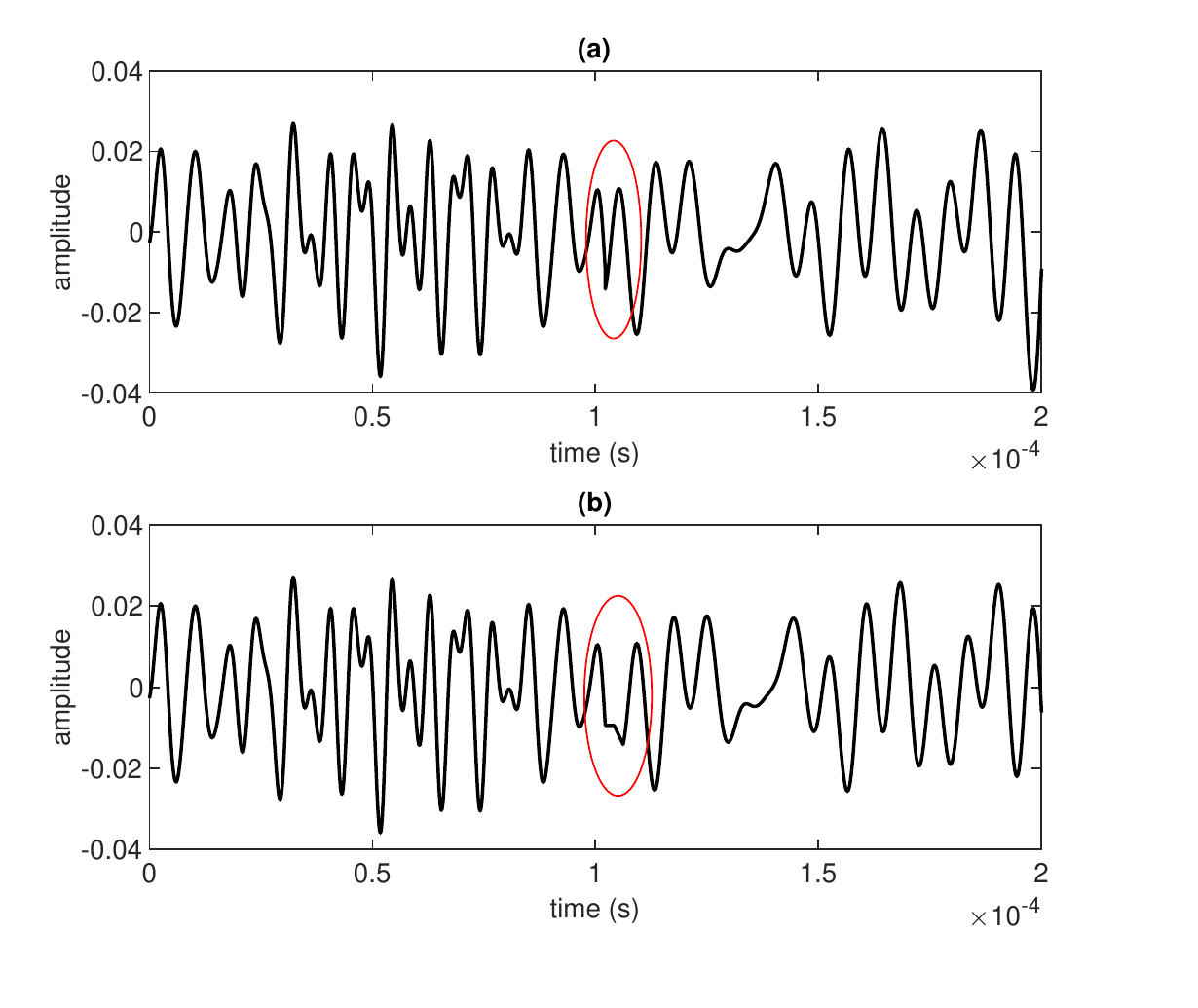}
    \vspace*{-8mm}
    \caption{Appended fast-time IF signals (a) before linear fitting (b) after linear fitting.}
    \label{fig:linear_fit}
\end{figure}

In order to exhibit the importance of this simple step, we simulate consecutive chirp response received by the radar from a target at a distance of $2$m. The appended signal $\widetilde{y}(nT_s)$ without and with interpolation is shown in Fig.~\ref{fig:linear_fit}. 
The distinct discontinuity observed in Fig.~\ref{fig:linear_fit}(a) is substituted by a line joining the two chirp responses in Fig.~\ref{fig:linear_fit}(b). It is important to note that a duration of $T_{\mathrm{idle}}$ is introduced by interpolation in Fig.~\ref{fig:linear_fit}(b) to ensure that the micro-motion parameters can be accurately estimated.

\begin{figure}[ht]
    \centering
    \includegraphics[width=9cm,height=7cm]{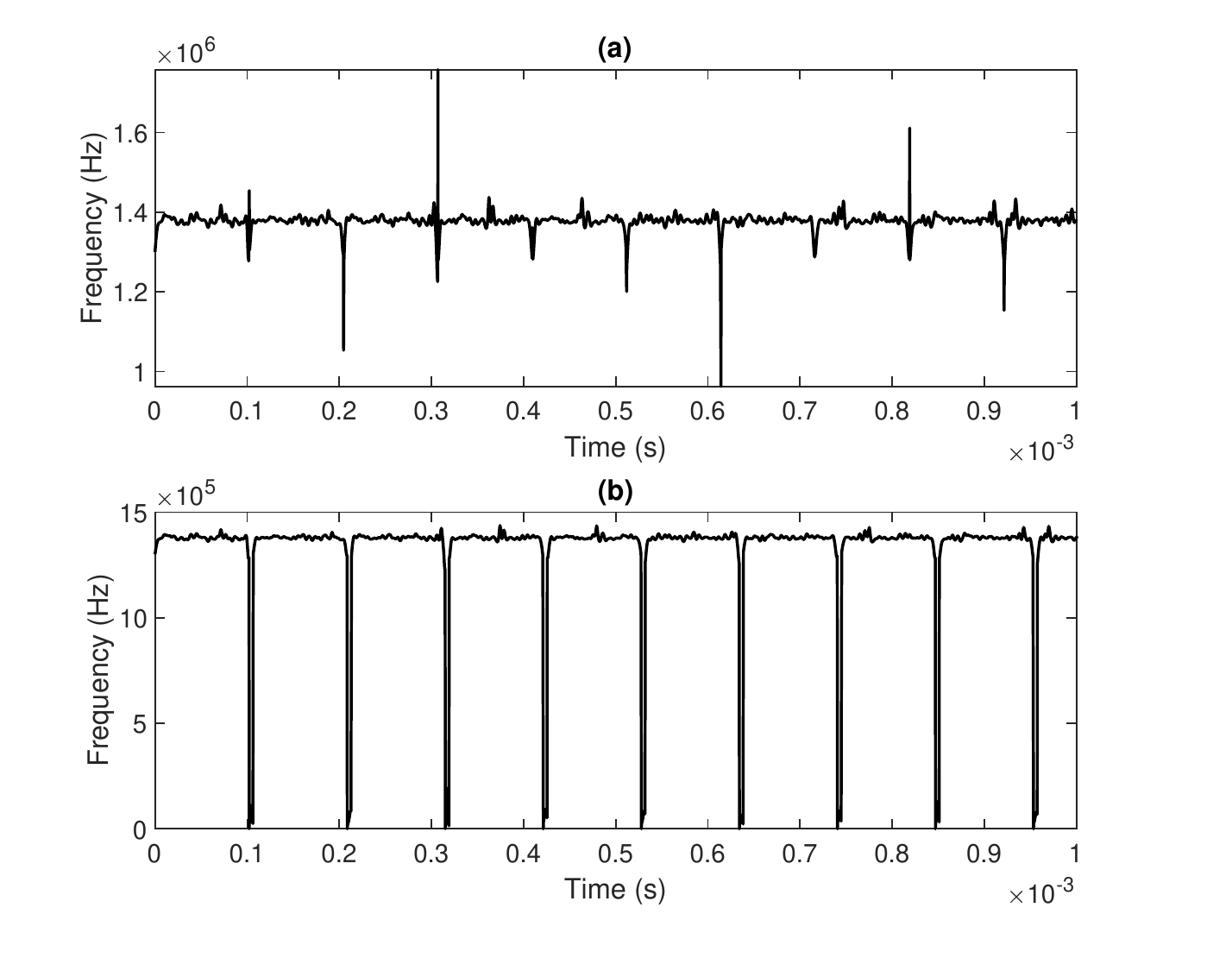}
    \vspace*{-8mm}
    \caption{Instantaneous frequency of fast-time IF signal (a) before linear fitting (b) after linear fitting.}
    \label{fig:spectrogram_linear_fit}
\end{figure}
The instantaneous frequency of the appended signal without and with interpolation between chirp responses are shown in Fig.~\ref{fig:spectrogram_linear_fit}. We observe an adhoc frequency variation after every chirp response in Fig.~\ref{fig:spectrogram_linear_fit}(a), while a low frequency component is observed in Fig.~\ref{fig:spectrogram_linear_fit}(b) due to the interpolation. Furthermore, the periodicity of the frequency change due to the idle time is different in the two cases because the former signal does not consider $T_{\mathrm{idle}}$.


\subsection{Empirical Mode Decomposition}

The response from various target parts that gives rise to micro-Doppler will be significantly lower than the body response. Furthermore, extraction of micro-Doppler signature in scenarios with low signal-to-noise ratio (SNR) will be difficult despite the use of RD filter. It is therefore necessary to extract the desired signature by exploiting signal characteristics. From the signal model, it was observed that the micro-Doppler signature is frequency modulated. We therefore consider the use of EMD to extract the desired signature.

EMD has been widely used in the literature to decompose various kinds of signals including signals that are non-linear, non-stationary and frequency modulated to mention a few ~\cite{2005Chengjie, 2015Brewster, 2011Yanbing, 2017Guo, 2008Bai, 2012Guanlei}. Any signal that is subjected to EMD is decomposed into several intrinsic mode functions (IMF) using Sifting process. IMFs are time-domain signals that satisfy two criteria: (1) Number of zero-crossings equal to the number of extrema, or differ by utmost one. (2) The mean of envelopes obtained by maximas and minimas should be equal to zero. An interesting characteristic of IMF is that the instantaneous frequency can change with time. Moreover, at any given time instant the instantaneous frequency in an IMF cannot be higher than that of the previous IMF. 

Any signal $x(n)$ that is subjected to EMD can be expressed as a sum of all the IMFs and the residue signal $r(n)$ as 
\begin{equation}
    x(n) = \sum_{i=1}^{I}c_{i}(n) +r(n),
\end{equation}
where $c_i(n)$ is the $i$th IMF. The decomposition is complete when the residue signal can no more be decomposed as another IMF. The Sifting process used to obtain IMFs is outlined in Algorithm~\ref{alg_emd}.
\RestyleAlgo{boxruled}
\LinesNumbered
\begin{algorithm}[ht]
  \caption{EMD \label{alg_emd}}
 Let $a_i(n)$ denote the residue signal obtained just after an IMF signal is extracted, and $b_{(j-1)}(n)$ be an intermediate signal used to evaluate an essential criterion for IMF determination. Initialize $i = 1$ and $a_{0}(n) = x(n)$.
    
Initialize $j = 1$ and $b_{0}(n) = a_{i-1}(n)$.

Construct an envelope signal $e_{max}(n)$ by interpolating all the maxima of $b_{j-1}(n)$ using a cubic spline. Likewise, the envelope $e_{min}(n)$ is obtained by interpolating all the minima of $b_{j-1}(n)$. Calculate the mean value between the envelopes $e_{max}(n)$ and $e_{min}(n)$ as $e_{mean}(n) = \frac{e_{max}(n) + e_{min}(n)}{2}$.


Calculate the difference signal $b_{j}(n) = b_{j-1}(n) - e_{mean}(n)$. 

If $b_{j}(n)$ meets the essential criteria to qualify as an IMF, then assign $c_{i}(n) = b_{j}(n)$ and move to Step 6. If the criterion is not met, increment the value of $j$, return to Step 3, and repeat the algorithm. Since accomplishing $e_{mean}(n)=0, \forall n$ is difficult, the criterion is relaxed to
    \begin{equation}
        \sum_{n=1}^{N}\frac{(b_{i(j-1)}(n)-b_{ij}(n))^2}{b_{ij}(n)^2} < \gamma,
    \end{equation}
where $\gamma$ is a predetermined threshold.

Assign the residual signal as $a_{i}(n) = a_{i-1}(n) - c_{i}(n)$.

If $a_{i}(n)$ has at-least the predetermined number of extrema, then increment the value of $i$ and continue to Step 2. Else stop the decomposition process.
\end{algorithm}

As IMFs are derived from the signal itself, they are adaptive to the intrinsic changes in the signal thereby preserving the micro-Doppler characteristics. It is also important to note that the spectral components of these IMFs vary from high to low gradually because the sifting process starts from the residual signal of the previous step. We can therefore visualize EMD as an adaptive filter bank that separates out various frequencies present in a signal.

Since the signal of interest $\widetilde{y}(nT_s) $ is complex in nature, we employ complex EMD (CEMD)~\cite{2007Rilling}. In CEMD, the envelope of complex-valued signal is considered as a $3$-dimensional tube that encloses the signal. The direction of this envelope can be defined as $\phi_k = 2{\pi}k/K, 1\leq{k}\leq{K}$, where $K$ denotes number of directions. For a complex-valued signal $x_c(n)$, IMF is generated using the same steps as that of EMD except for the intermediate signal generated during sifting process. The algorithm for generating the intermediate signal in CEMD is summarized in Algorithm~\ref{alg_cemd}.
\RestyleAlgo{boxruled}
\LinesNumbered
\begin{algorithm}[ht]
  \caption{Intermediate signal computation for CEMD \label{alg_cemd}}
For the $i$th IMF extraction, initialize $b_0(n) = a_{i-1}(n)$. For $1\leq{k}\leq{K}$, perform the steps 2 and 3.

Project $b_{j-1}(n)$ onto the direction $\phi_k$ such that $P_{\phi_k}(n) = \mathrm{Re}(e^{(-j\phi_k)}b_{j-1}(n))$, where Re(.) denotes real part of complex-valued signal within.
    
Interpolate ${(d_j^k,b_{j-1}(d_j^k))}$ where $d_j^k$ is the location of the maxima of $P_{\phi_k}(n)$. This creates an envelope $e_{\phi_k}(n)$ of $b_{j-1}(n)$ in direction $\phi_k$. After the envelopes are generated for all directions, perform steps 4 and 5.

Compute the mean such that $e_{mean}(n) = (1/K)\sum_{k=1}^{K}e_{\phi_k}(n)$

Subtract the mean $e_{mean}(n)$ from $x_c(n)$ to obtain the intermediate signal $b_j(n) = b_{j-1}(n) - e_{mean}(n)$.
\end{algorithm}

The $I$ complex-valued IMFs obtained by CEMD have similar characteristics as that of their real counterparts obtained using EMD. Unlike the EMD of a real signal where the instantaneous frequency is estimated from the Hilbert-Huang transform of the IMF, instantaneous frequency of the $i$th complex IMFs can be obtained directly from the instantaneous phase $\varphi(nT_s)$ of $c^{(i)}(n)$ as
\begin{equation}
f^{(i)}_{\mathrm{inst}}(n) = \frac{1}{2\pi} \frac{d\varphi(nT_s)}{dt}.
\end{equation}
Since $f^{(i)}_{\mathrm{inst}}(n)$ directly relates to the rotation rate of the IMF phasor, it satisfies the condition
\begin{equation}
f^{(i)}_{\mathrm{inst}}(n)>f^{(i+1)}_{\mathrm{inst}}(n).
\end{equation}

In the proposed context, the discontinuity between the consecutive chirps, when appended without linear fitting, introduces an arbitrary change in the instantaneous frequency of each IMF similar to what was observed in Fig.~\ref{fig:spectrogram_linear_fit}(a).  This discontinuity adversely affects the micro-Doppler signature analysis. The use of linear fitting on the contrary, will introduce a consistent low frequency component over the segment $\mathbf{y}_{\mathrm{interp}}^{(l,l+1)}$. Although the micro-Doppler information is not available over this segment, the time-frequency analysis is not impacted by the impulsive noise due to the discontinuity. 

\subsection{IMF dimensionality reduction}
Since $\widetilde{y}(nT_s)$ is obtained after RD filtering, the additive noise outside of its passband is suppressed. The use of CEMD is expected to segregate the frequency-modulated micro-Doppler signature into a subset of the IMFs while the in-band noise decomposed into other IMFs. We therefore reduce the the number of IMFs used to reconstruct a high fidelity signal $\widehat{y}(nT_s)$ as
\begin{align}
\widetilde{y}(nT_s) &= \widehat{y}(nT_s)+r_{\mathrm{inBand}}(nT_s) & &\\
&=\sum_{i=1}^{I_r}c^{(i)}(nT_s) + r_{\mathrm{inBand}}(nT_s), & &
\end{align}
where $I_r<I$ is the reduced number of IMFs employed for reconstruction and $c^{(i)}(nT_s)$ are those IMFs having target signature. $r_{\mathrm{inBand}}(nT_s)$ includes the remaining IMFs and the residue. Identification of the $I_r$ IMFs that preserve target signature is critical for this dimensionality reduction. 

Towards that, we first note $\widetilde{y}(nT_s)$ to be centered around the range frequency $f_{R_0}$. With this range frequency known, we can infer that the IMFs with instantaneous frequency that is close to $f_{R_0}$ preserve target signature. However, there are two concerns to determine the proximity of IMF instantaneous frequency with $f_{R_0}$. As is well known, the instantaneous frequency of an IMF can change with time. While this asserts that the IMF can preserve frequency modulated micro-Doppler signature, the proximity of instantaneous frequency to $f_{R_0}$ over time has to be observed to determine if the IMF indeed has the desired signature. Furthermore, the instantaneous frequency over the segments $\mathbf{y}_{\mathrm{interp}}^{(l,l+1)}$ will be low in all the IMFs. One has to exclude the instantaneous frequency over these segments while evaluating the proximity of the IMF instantaneous frequency with the target range.

Before evaluating the instantaneous frequency of $c^{(i)}(n)$, we also exclude those time instances where the the magnitude of the IMF is close to the zero crossing. The criteria for determining the proximity of the IMF to the target range frequency are
\begin{align}
\label{Crit1}
d^{(i)} &= |\mathrm{mean}(f^{(i)}_{\mathrm{inst}}(n))-\hat{f}_{R_0}| < \Delta f_{\mathrm{mD}}/2,& &\\
\label{Crit2}
\sigma (f^{(i)}_{\mathrm{inst}})&<\Delta f_{\mathrm{mD}}/2& &
\end{align}
where $\Delta f_{\mathrm{mD}}$ is the two-sided maximum anticipated micro-Doppler frequency spread given in (\ref{maxMD-Spread}) and $\sigma \big(f^{(i)}_{\mathrm{inst}}\big)$ is the standard deviation of $f^{(i)}_{\mathrm{inst}}(n)$. As the target signature is centered around $\hat{f}_{R_0}$, it is observed that the first few complex-valued IMFs preserve the target signature. This encourages us to restrict to the complex EMD only up to the first $I_r$ IMFs, thereby reducing the computational complexity. The study of the number of IMFs required for reconstruction under various conditions is presented in the next section.

\section{Simulation Studies}
\begin{figure}[ht]
    \centering
     \includegraphics[width=9cm, height=7cm]{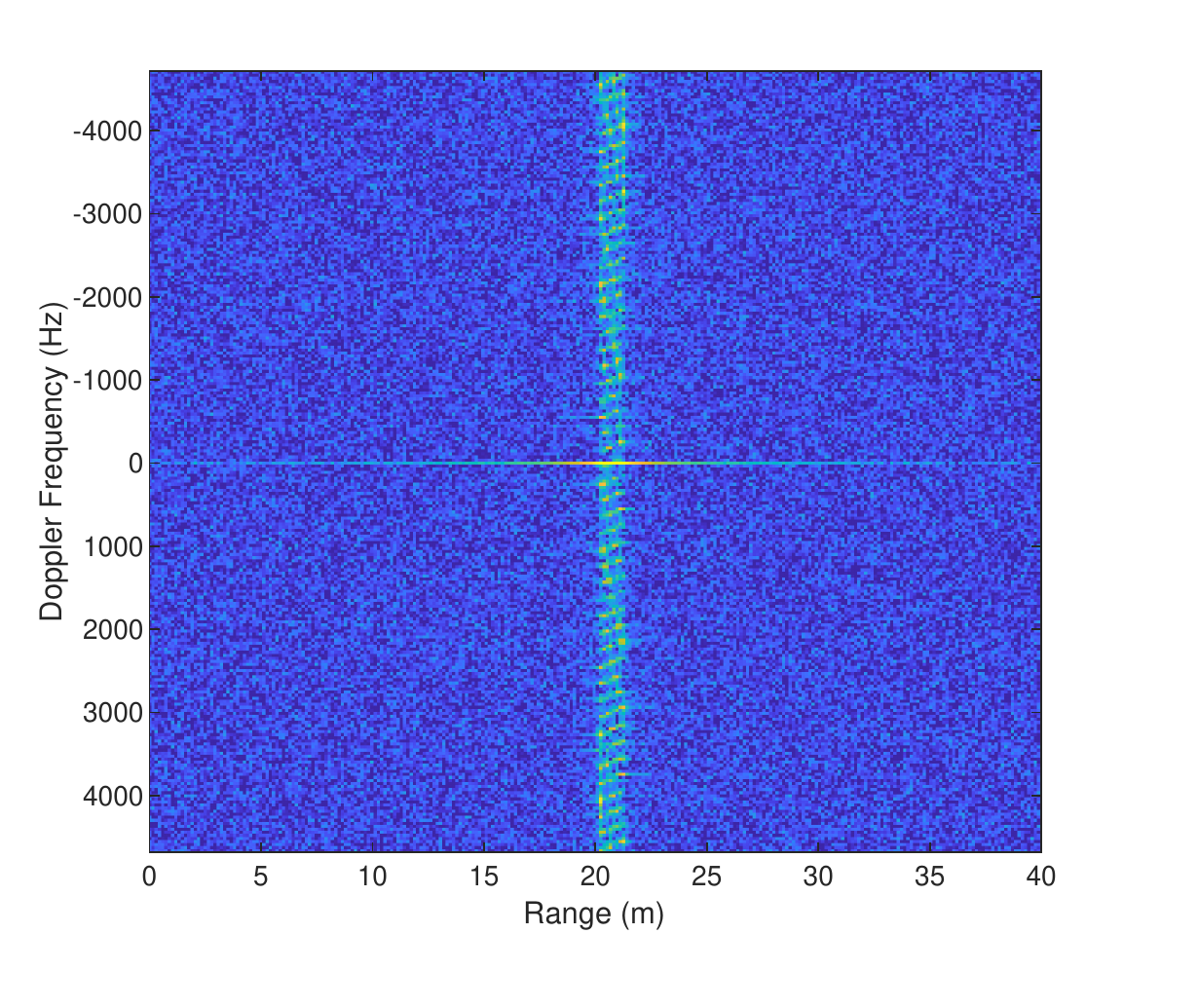}
    \vspace*{-8mm}
    \caption{RD map for the scene under consideration.}
    \label{fig:RDmapIllustr}
 \end{figure}

We first illustrate the capability of the proposed technique to extract the high frequency micro-Doppler signature. A target with $3$ blades rotating at $12000$ rpm is considered, at a distance of $20$m from the radar. Table~\ref{simconfig} provides the radar configuration used for this study.
\\
\begin{table}[ht]
\begin{tabular}{ |p{4cm}|p{3cm}| }
 \hline
 \multicolumn{2}{|c|}{Radar parameters} \\
 \hline
Start frequency, $f_0$   & $77GHz$ \\
 \hline
 Chirp rate, $\mu$   & $10MHz/\mu{s}$ \\
 \hline
 Chirp duration, $T_c$ & $102.4\mu{s}$\\
 \hline
 Chirp repetition interval, $T_{\mathrm{CRI}}$ & $104.43\mu{s}$\\
 \hline
 Sampling rate, $f_s$ & $5MHz$\\
 \hline
 Bandwidth, $B$ &   $1.02GHz$ \\
 \hline
 Number of Chirps within CPI & $256$ \\ \hline
\end{tabular}\\
     \caption{Radar configuration for the simulation.}
     \vspace{-4mm}
    \label{simconfig}
\end{table}
The signal is synthesized using (\ref{SM_md})-(\ref{IF_sig_one_blade}) with $5~$dB SNR, and the response due to the blade is considered to be $6~$dB lower than the body response $\alpha^{body}$. The RD map obtained using $256$ chirp responses is shown in Fig.~\ref{fig:RDmapIllustr}.
We observe that the target response is spread over the entire Doppler range asserting a severely aliased case. A spread along the range dimension is also observed because $\Delta f_{\mathrm{mD}}>c/(2B)$.

\begin{figure}[H]
    \centering
    \includegraphics[width=\linewidth]{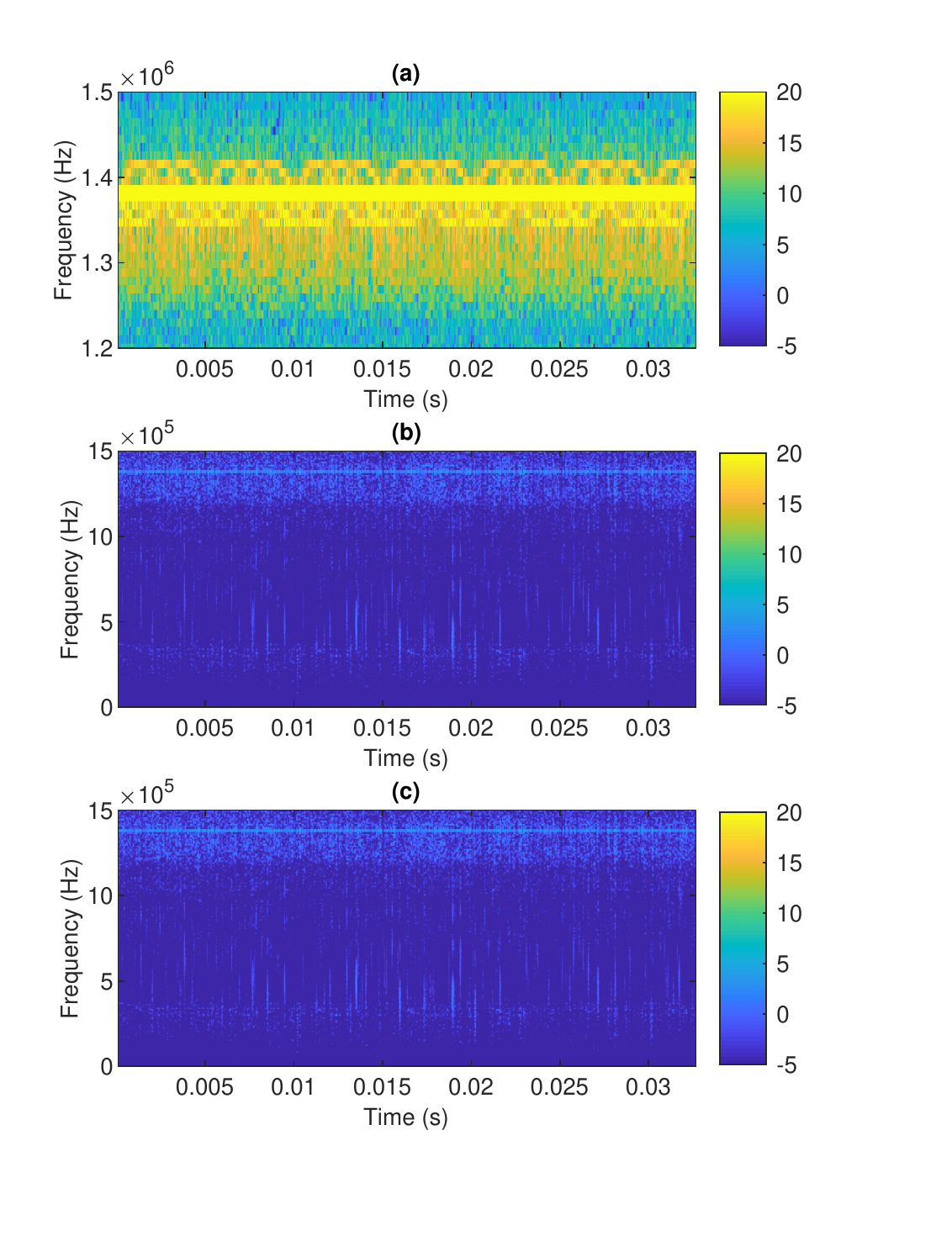}
    \vspace*{-15mm}
    \caption{Spectrogram when the SNR is 5dB for (a) IMF 1 (b) IMF 2 and (c) IMF 3.}
    \label{fig:imf1_imf2_5db}
\end{figure}

The signal is subjected to RD filtering with the range and Doppler filter bandwidths set to the maximum micro-Doppler frequency spread and $f_{\mathrm{crf}}$, respectively. The reconstructed $y_{\mathrm{filt}}^{(l)}(nT_s)$ is concatenated with linear interpolation before CEMD. Time-frequency analysis of the IMFs is performed using spectrogram with a window size of $102.4~\mu$s with $90\%$ overlap. We inspect the spectrogram of the first three IMFs for the micro-Doppler signature in Fig.~\ref{fig:imf1_imf2_5db}.
The spectrogram for the first IMF shows three sinusoids corresponding to the three rotating blades of the target. This sinusoidal frequency variation is centered around the target range frequency along with the response due to the target body. While the out-of-band noise suppression is accomplished predominantly by the RD filter, the in-band noise is reduced by the use of CEMD. The spectrograms of the other two IMFs exhibit weak in-band noise that can be discarded.

\begin{figure}[ht]
    \centering
    \vspace{-3mm}
    \includegraphics[width=9cm, height = 7cm]{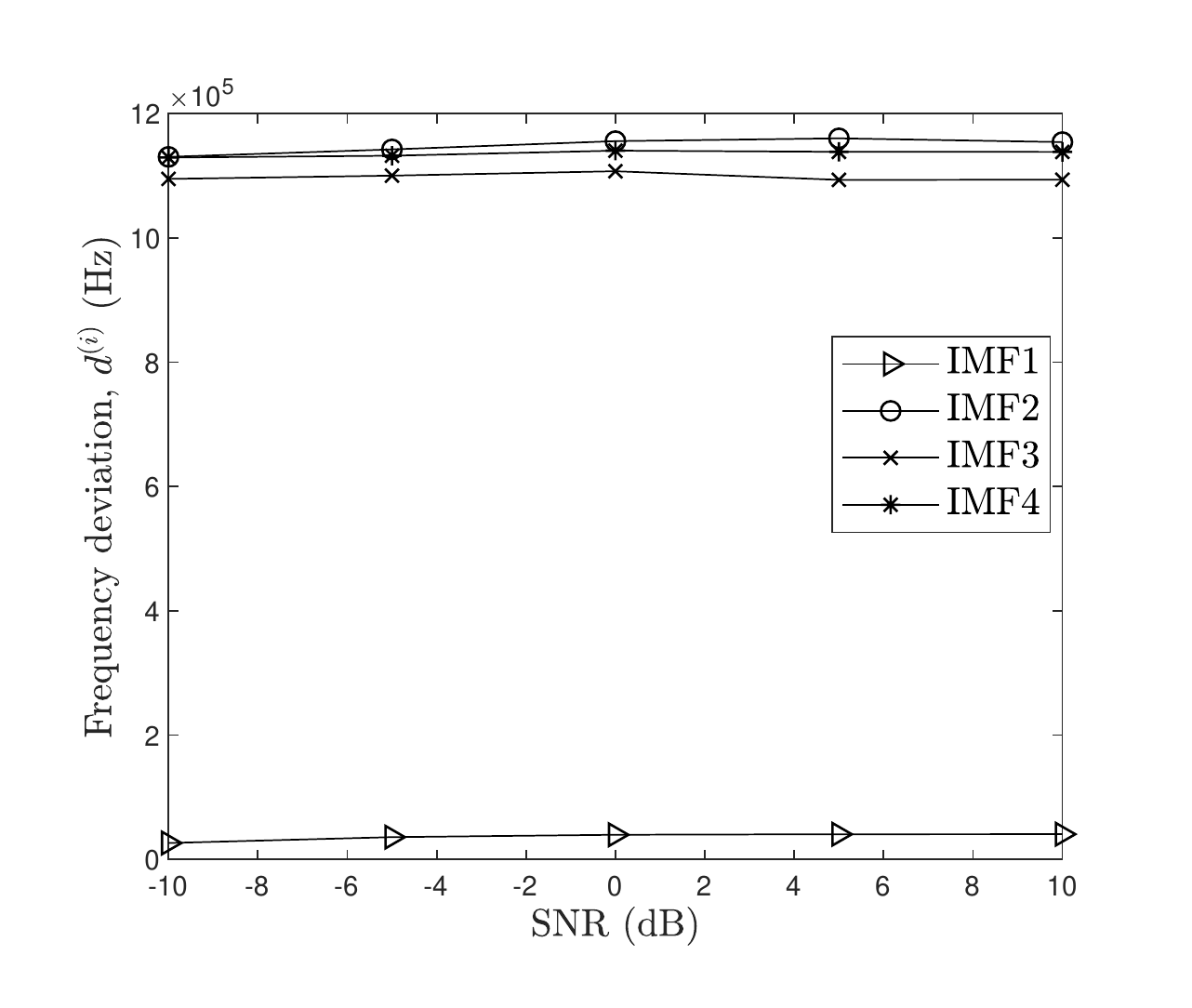}
    \vspace{-3mm}
    \caption{Performance of $d^{(i)}$ against SNR for the first four IMFs.}
   \vspace{-3mm}
    \label{fig:-FreqDevVsSNR}
 \end{figure}

\begin{figure}[ht]
    \centering
    \vspace{-3mm}
    \includegraphics[width=9cm, height = 7cm]{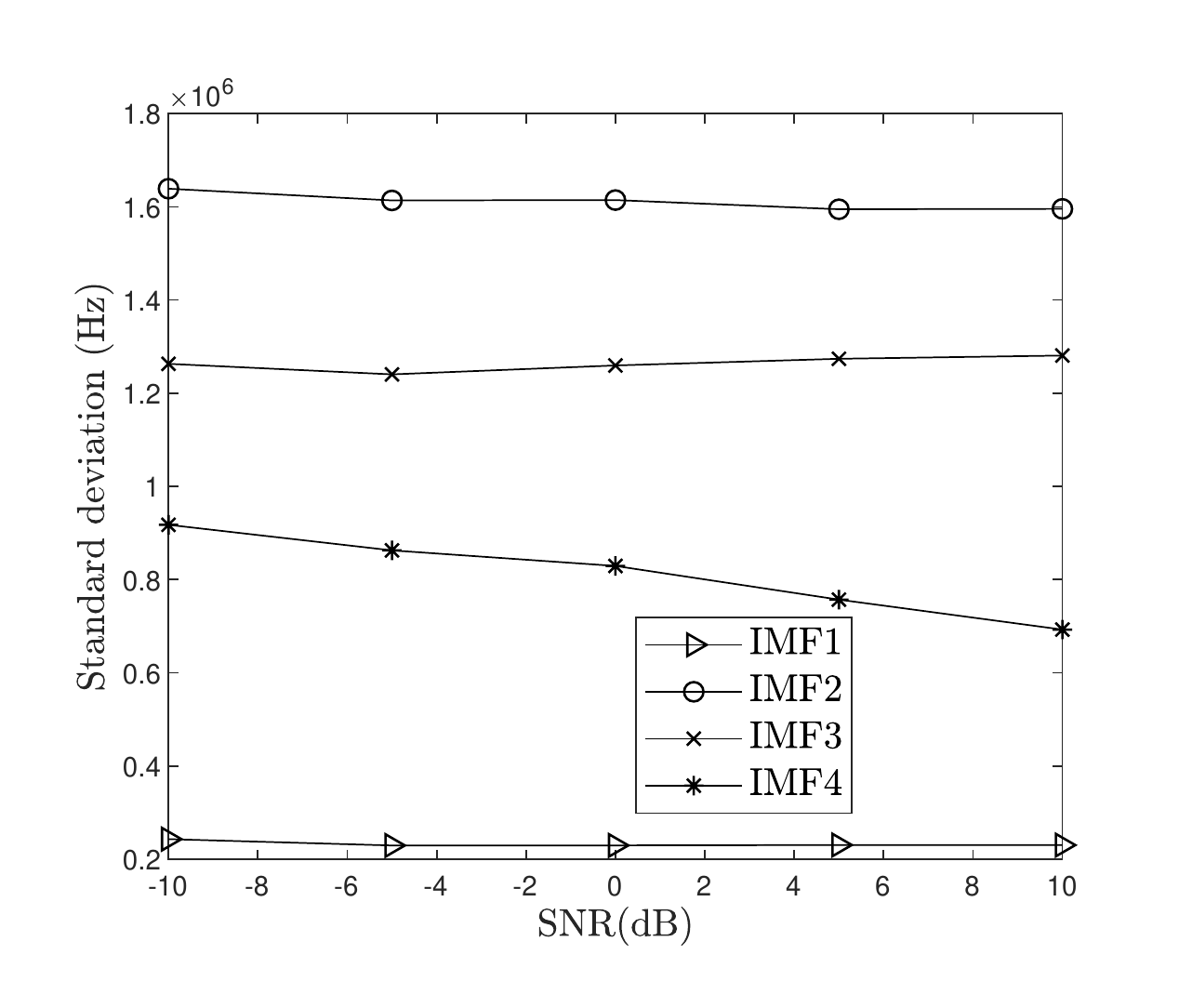}
    \vspace{-7mm}
    \caption{Performance of $\sigma(f^{(i)}_{\mathrm{inst}})$ against SNR for the first four IMFs.}
    \vspace{-3mm}
    \label{fig:-StdDevVsSNR}
   \end{figure}

The criteria presented in (\ref{Crit1}) and (\ref{Crit2}) enables one to automatically choose the IMFs required to reconstruct the signal. This choice is particularly important in minimizing the in-band noise without compromising the micro-Doppler signature. In order to verify the efficacy of these criteria, we perform Monte-Carlo simulation over 100 trials in which the frequency deviation $d^{(i)}$ and the standard deviation $\sigma(f^{(i)}_{\mathrm{inst}})$ are studied against SNR. Figures~\ref{fig:-FreqDevVsSNR} and \ref{fig:-StdDevVsSNR} show the performance of these metrics for the first four IMFs against SNR.

From Fig.~\ref{fig:-FreqDevVsSNR}, we note that $d^{(i)}$ is the least for the first IMF at all SNR conditions. This indicates that the instantaneous frequency of first IMF is the nearest to the target range frequency compared to all other IMFs. The standard deviation $\sigma(f^{(i)}_{\mathrm{inst}})$ for the first IMF is also observed in Fig.~\ref{fig:-StdDevVsSNR} to be the least compared to other IMFs, thereby indicating that the micro-Doppler signature is most often embedded within the first IMF. This inference is particularly important because one needs to only compute the first IMF for the complex signal. This greatly reduces the real-time computational complexity.

It is also important to note from Fig.~\ref{fig:-StdDevVsSNR} that the standard deviation of IMF 2 is greater than that of IMFs $3$ and $4$. When the target signature that is centered around $f_{R_0}$ is embedded in the first IMF, the in-band noise will be pushed to the second IMF. The high frequency variations in the noise therefore exhibits a large $\sigma(f^{(i)}_{\mathrm{inst}})$ for the second IMF. The residual noise components are distributed to third and higher IMFs which show reduced variance.

\begin{figure}[ht]
\vspace{-5mm}
    \centering
    \includegraphics[width=\linewidth]{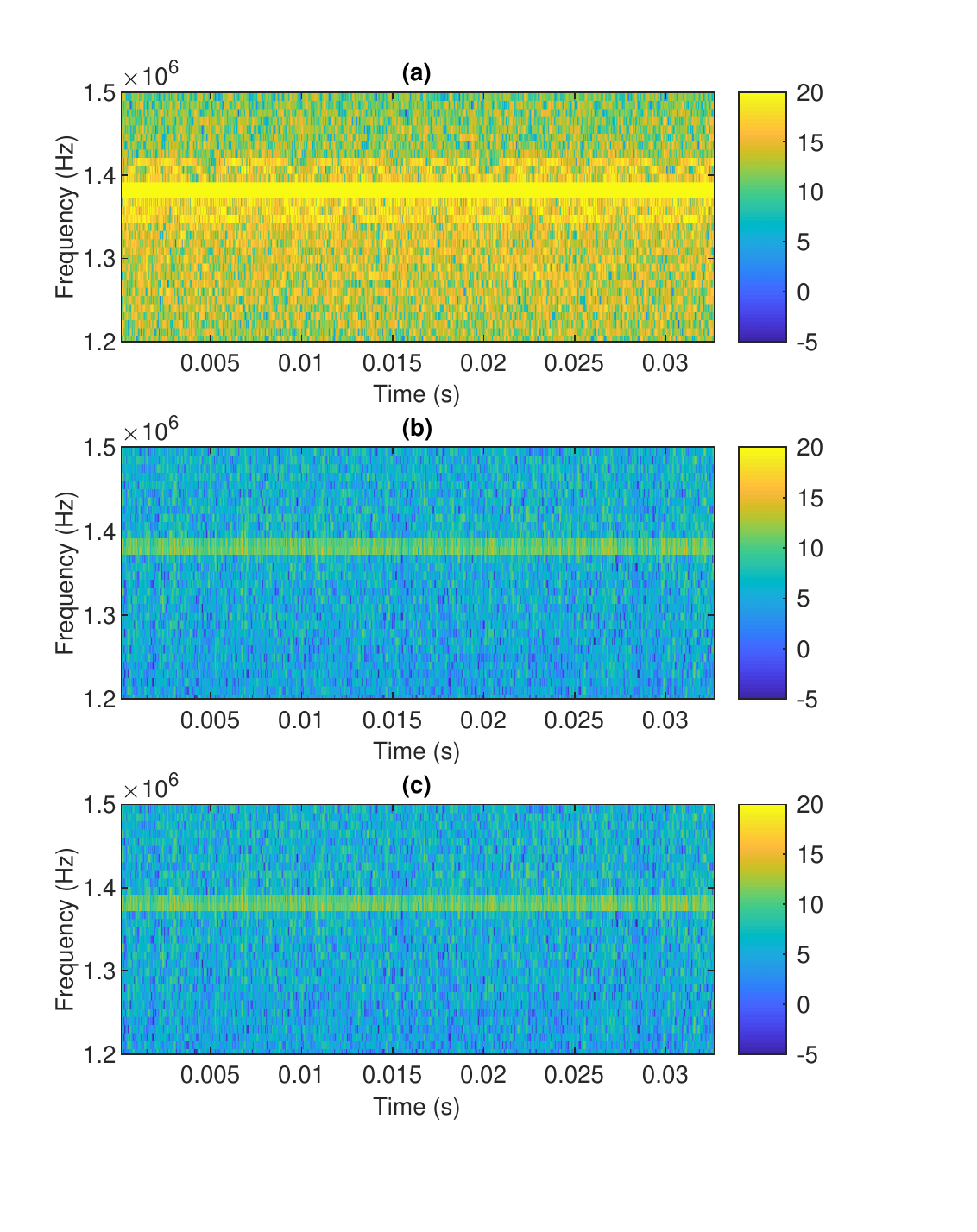}
    \vspace*{-15mm}
    \caption{Spectrogram when the SNR is $-5dB$ for (a) IMF 1 (b) IMF 2 and (c) IMF 3.}
    \label{fig:imf1_imf2_-5db}
\end{figure}

Despite Fig.~\ref{fig:-FreqDevVsSNR} suggesting the use of only the first IMF for micro-Doppler signature analysis, we validate the same for a scene with $-5~$dB SNR. The spectrogram obtained for the first three IMFs is shown in Fig.~\ref{fig:imf1_imf2_-5db}.

We note that the first IMF continues to have the micro-Doppler signature despite having some noise compared to the case with $5~$dB SNR observed in Fig.~\ref{fig:imf1_imf2_5db}. The second and third IMFs are observed to have residual body response and noise.

\section{Experimental studies}
In this section, we present the extraction of micro-Doppler signature from UAV using $77~$GHz TI AWR1642BOOST-ODS radar. The data is acquired indoors in a controlled environment with a Quanser drone hovering at a distance of $3~$m from the radar and $1~$m above the ground. The UAV has four rotors with two blades, each of length $6~$cm. The blades rotate at unknown speed that is necessary for the drone to hover at one position. The raw ADC data acquired using the DCA1000 EVM card is post processed using the proposed algorithm. Table.~\ref{exp3config} provides the radar configuration details used for the experiment.
\\
\begin{table}[ht]
\begin{tabular}{ |p{4cm}|p{3cm}| }
 \hline
 \multicolumn{2}{|c|}{Radar parameters} \\
 \hline
 Chirp rate, $\mu$   & $25MHz/\mu{s}$ \\
 \hline
 Chirp duration, $T_c$ & $102.4\mu{s}$\\
 \hline
 Chirp repetition interval, $T_{\mathrm{CRI}}$ & $114.43\mu{s}$\\
 \hline
 Sampling rate, $f_s$ & $5MHz$\\
 \hline
 Bandwidth, $B$ & $2.5GHz$ \\
 \hline
 Total chirps & $625$  \\
 \hline
\end{tabular}\\
     \caption{Radar Configuration for the experiment.}
    \label{exp3config}
\end{table}

\begin{figure}[ht]
\vspace{-5mm}
    \centering
    \includegraphics[width=\linewidth, height = 4.5cm]{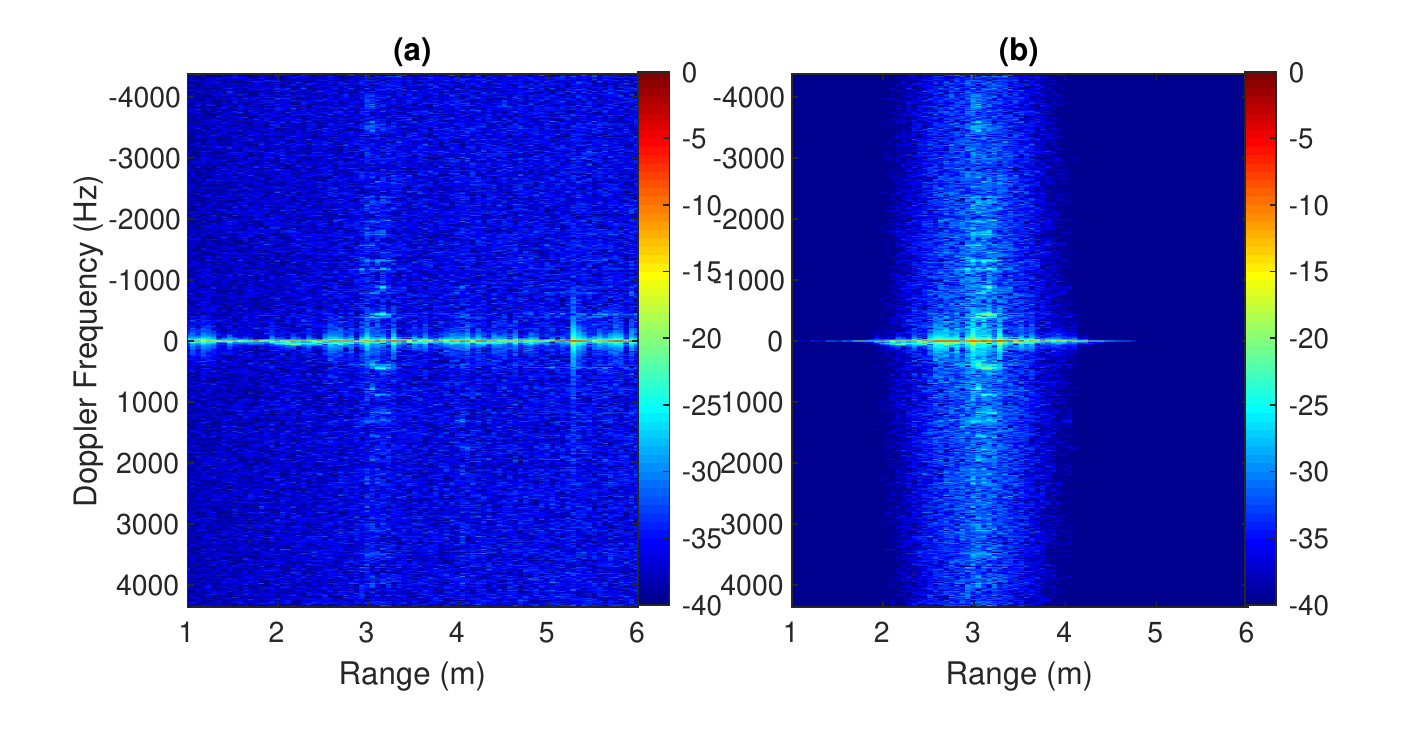}
    \vspace{-3mm}
    \caption{RD map for the experiment (a) before filtering, (b) after filtering.}
    \label{fig:RD_maps}
\end{figure}

To extract the micro-Doppler signature of drone we process the radar data using the conventional STMDSE approach and the proposed approach. For the proposed approach, we use all the $625$ chirps to generate the RD map as shown in Fig.~\ref{fig:RD_maps}(a). The RD map obtained after RD filtered with a range cut-off frequency $f_c = 168.6 kHz$ is shown in Fig.~\ref{fig:RD_maps}(b). For STMDSE and the proposed technique, we assume the target range is known apriori. 

\begin{figure}[H]
    \centering
    \includegraphics[width=\linewidth, height = 4.5cm]{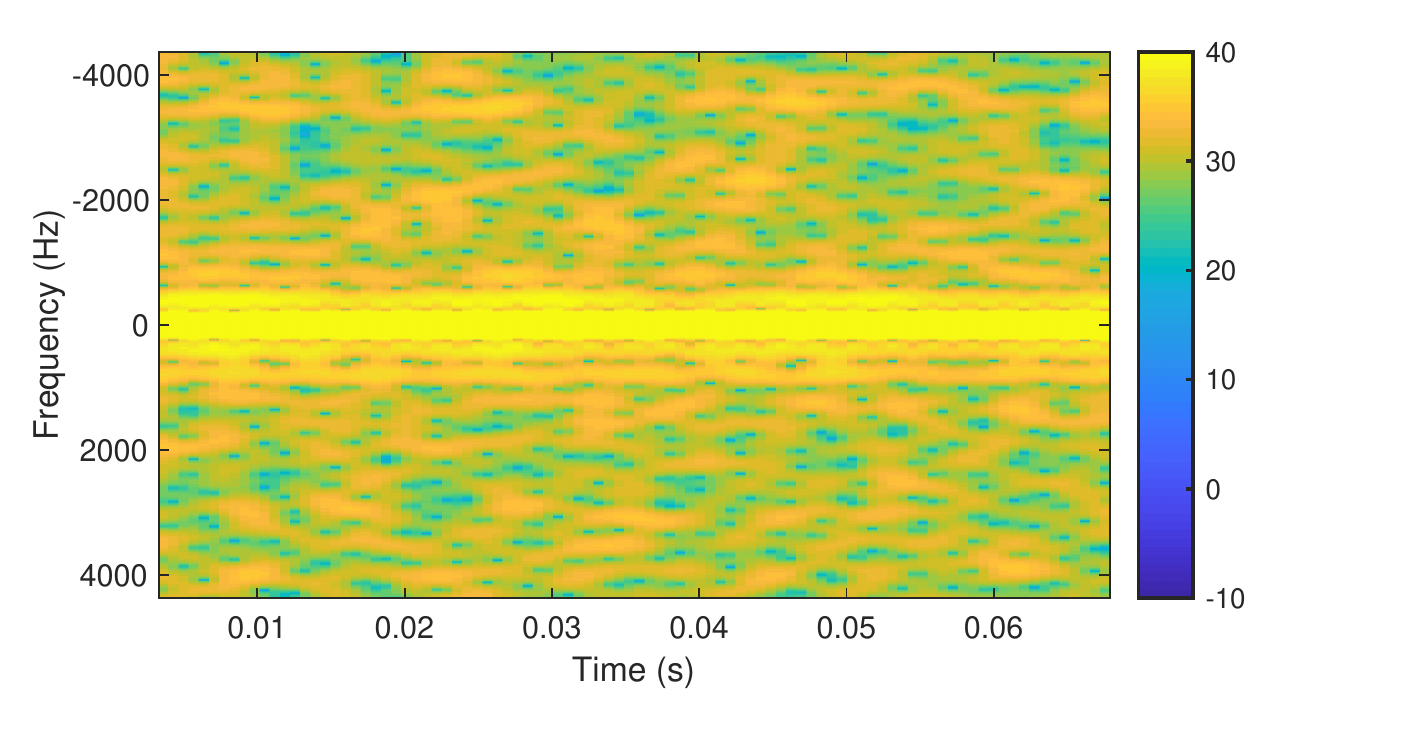}
    \caption{Spectrogram of the UAV obtained by STMDSE.}
    \label{fig:exp_after_con_approach}
\end{figure}

The spectrogram for STMDSE is shown in Fig.~\ref{fig:exp_after_con_approach}.
It is observed that the micro-Doppler signature does not exhibit any specific pattern in the spectrogram. This is due to the aliasing effect introduced by the micro-Doppler frequency which varies beyond $f_{\mathrm{crf}}$. We can therefore conclude that STMDSE is not suitable for such applications.

\begin{figure}[H]
    \centering
    \includegraphics[width=\linewidth, height=4.5cm]{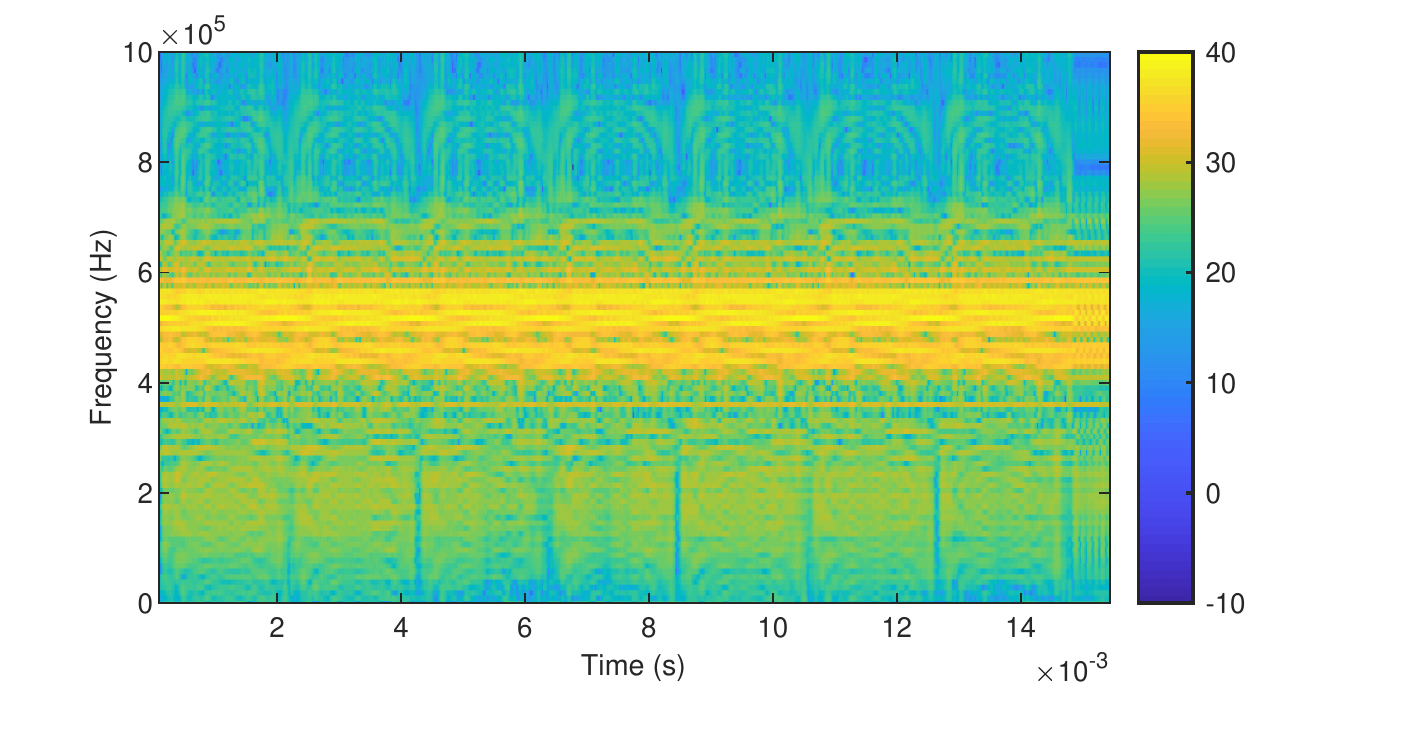}
    \caption{Spectrogram of the UAV obtained by the proposed approach.}
    \label{fig:exp_after_emd}
\end{figure}

The signal $\widehat{y}(nT_s)$ is reconstructed using around $152$ chirp responses instead of all the $625$ chirps. With $d^{(1)}=22.7 kHz$ and $\sigma(f^{(1)}_{\mathrm{inst}})=121.5 kHz$, $d^{(2)}=338.3 kHz$ and $\sigma(f^{(2)}_{\mathrm{inst}})=209.42 kHz$, only the first IMF is selected to reconstruct the signal. The resultant micro-Doppler signature obtained from the spectrogram is shown in Fig.~\ref{fig:exp_after_emd}.
We observe a frequency spread of about $80~$kHz around the target range frequency of about $5.4~$MHz in the spectrogram. This is comparable with the previous simulation, although the periodicity is not clearly visible due to the varying rotor speed and the random initial phase of each blade. Comparing the results obtained in Fig. ~\ref{fig:exp_after_con_approach} and Fig.~\ref{fig:exp_after_emd}, it is evident that the proposed approach can effectively extract high-frequency micro-Doppler signature which is otherwise not possible to obtain using the conventional approach. 

\section{Conclusion}
In this work we presented an EMD-based approach for high frequency micro-Doppler signature extracted from the fast time FMCW radar acquisitions. Due to the high sampling rate along the fast time, it was shown that the micro-Doppler signature can be extracted without aliasing with this approach. The proposed technique is shown to address the prevalent discontinuities between chirp responses. The use of CEMD enabled the segregation of micro-Doppler signature from the in-band noise before the time-frequency analysis of the signal reconstructed from selected IMFs. Besides verifying the extraction of the micro-Doppler signature, simulation studies also verified the efficacy of the criteria proposed for IMF selection. Experimental results for a hovering UAV verified the use of proposed approach for applications that encounter high frequency micro-motions.

%

\begin{IEEEbiography}{Michael Shell}
Biography text here.
\end{IEEEbiography}

\begin{IEEEbiographynophoto}{John Doe}
Biography text here.
\end{IEEEbiographynophoto}


\begin{IEEEbiographynophoto}{Jane Doe}
Biography text here.
\end{IEEEbiographynophoto}




\end{document}